\documentclass[reqno,12pt]{amsart}
 \usepackage{amssymb, latexsym, euscript}
\usepackage{amssymb}
\usepackage{amscd}
\usepackage{graphicx}
\usepackage{amstext}
\usepackage{amsfonts}
\usepackage{tikz}
\usepackage{amsbsy}
\usetikzlibrary{arrows,decorations.pathmorphing,backgrounds,positioning,fit}
\newtheorem{theorem}{Theorem}[section]

 \newtheorem{lemma}[theorem]{Lemma}
 
 \newtheorem{corollary}[theorem]{Corollary}

\theoremstyle{definition}
\newtheorem{example}[theorem]{Example}

\numberwithin{equation}{section}

\begin{document}
\title[Solvable models
with nonlocal one point interactions]{Quantum solvable models with
nonlocal one point interactions} \author[S.~Kuzhel]{Sergii Kuzhel}
\author[M.~Znojil]{Miloslav Znojil}

\address{AGH University of Science and Technology \\ 30-059 Krak\'{o}w, Poland}
\email{kuzhel@agh.edu.pl}

\address{Nuclear Physics Institute ASCR, Hlavn\'{\i} 130, 250 68 \v{R}e\v{z}, Czech
Republic} \email{znojil@ujf.cas.cz}


\maketitle

\begin{abstract}
Within the framework of quantum mechanics working with
one-dimensional, manifestly non-Hermitian Hamiltonians $H=-{d^2}/{dx^2}+V$ the
traditional class of the exactly solvable models with local point
interactions $V=V(x)$ is generalized and studied. The consequences of the use of
the nonlocal point interactions such that $(V f)(x) = \int K(x,s)
f(s) ds$ are discussed using the suitably adapted formalism of
boundary triplets.
\end{abstract}

\section{Introduction}\label{intro}
The authors of introductory textbooks on Quantum Mechanics have to
combine a persuasive survey of its heuristics (involving, e.g., the
explanation of the so called principle of correspondence) and
applicability (say, to hydrogen atom) with a maximally compact
presentation of the underlying mathematics. This means that a more
advanced  understanding of the theory proceeds, typically, either
beyond the naive forms of the classical-quantum correspondence, or
beyond the oversimplified usage of the underlying language of
functional analysis.

Both of these tendencies appeared re-unified after the mind-boggling
discovery \cite{CGM} - \cite{BB} of the existence of certain rather anomalous
one-dimensional Schr\"{o}dinger operators
 \begin{equation}\label{selok}
  H=-\frac{d^2}{dx^2}+ V(x)
 \end{equation}
in Hilbert space $L_2(\mathbb{R})$ which appeared to possess real
spectra and to support stable bound states in spite of being {\em
manifestly non-self-adjoint}.

The existence of such an apparent puzzle encouraged an
intensification of the study of similar non-self-adjoint operators
which led, recently, to its more or less satisfactory clarification
(cf., e.g., the mathematically oriented collection of reviews
\cite{book} of the current situation in the field). {\it A priori},
it is not too surprising that the reliable physical interpretation
of the manifestly non-self-adjoint bound-state models (\ref{selok})
may prove mathematically deeply nontrivial \cite{Dieudonne}. 

In the context of the non-self-adjoint-operator phenomenology 
serious difficulties emerged in the scattering dynamical regime
\cite{Jones}. In this regime the (naturally, highly desirable!)
unitarity of the evolution can only be guaranteed {\em after} a
replacement of the local forces in (\ref{selok}) by their suitable
{\em non-local-interaction} generalizations \cite{scatt}
  $$
  V(x)f(x) \ \to \ \int_{-\infty}^\infty K(x,s) f(s) ds\,.
  $$
In such a situation one is exposed to the necessity of a {\em
simultaneous, viz., non-self-adjoint and nonlocal} generalization of
interactions.

In the present paper we study \emph{non-self-adjoint}
Schr\"{o}dinger operators with \emph{nonlocal one point
interactions}. Such kind of new solvable models with point
interactions has recently been proposed and studied (for
self-adjoint case) by S. Albeverio and L. Nizhnik \cite{Nizhnik} 
(see also \cite{Nizhnik1} - \cite{Nizhnik4}).
Our interest to the non-self-adjoint case was inspirited in part by
an intensive development of Pseudo-Hermitian
($\mathcal{PT}$-Symmetric) Quantum Mechanics PHQM  (PTQM) during
last decades \cite{Carl}--\cite{ZN}.

Non-self-adjoint point-interaction solvable models (see, e.g.,
\cite{nonlocal} -- \cite{ZN3}) require more detailed analysis in comparison with
theirs self-adjoint counterparts. In contrast to the self-adjoint
case \cite{AK_Albeverio0}, one should illustrate a typical PHQM/PTQM evolution of
spectral properties which can be obtained by changing parameters of
the model: complex eigenvalues $\to$ spectral singularities /
exceptional points $\to$  similarity to a self-adjoint operator. One
of the simplest examples of such kind is the well-studied
$\delta$-interaction model $-{d^2}/{dx^2}+a<\delta,\cdot>\delta(x)$
with complex parameter $a\in\mathbb{C}$ (see \cite{deltainteraction}, \cite{Grod}
or section \ref{subsec4.3} below). However, this model seems to be
sufficiently trivial due to the very simple structure of the
singular potential that leads to `poor' spectral properties of the
corresponding operator-realizations $H_a$ (for instance, $H_a$ have
no exceptional points and bound states on continuous spectrum).

One of possible `reasonable complication' of the model consists in
the addition of the nonlocal interaction term
$\int_{-\infty}^{\infty}K(x,s)f(s)ds$. Trying to keep the
solvability of the model and its intimate relationship with
$\delta$-interaction, we assume that
$$
K(x,s)=q(x)\delta(s)+\delta(x)q^*(s),
$$
where
$q\in{L_2(\mathbb{R})}$ is a given piecewise continuous function.
The corresponding nonlocal $\delta$-interaction
\begin{equation}\label{new1}
-\frac{d^2}{dx^2}+a<\delta,\cdot>\delta(x)+<\delta,\cdot>q(x)+(q,\cdot)\delta(x),
\quad a\in\mathbb{C},
\end{equation}
where $(\cdot,\cdot)$ is the inner product in $L_2(\mathbb{R})$ linear in the second argument,
is studied in Section \ref{sec4} with the use of boundary triplet technique (see the Appendix).
Namely, the formal expression (\ref{new1}) gives rise to the family of operators $\{H_a\}$:
$$
H_af=-\frac{d^2f}{dx^2}+f(0)q(x), \qquad a\in\mathbb{C}, \quad
q\in{L_2(\mathbb{R})} \quad \mbox{is fixed}
$$
with domains of definition (\ref{ggg1}) which are determined by the
singular part of perturbation
$a<\delta,\cdot>\delta(x)+(q,\cdot)\delta(x)$ in (\ref{new1}). Our
investigation of $\{H_a\}$ is based on the fact that each operator
$H_a$ is the proper extension of the symmetric operator
$\widetilde{S}_{min}$ (\ref{new131}), i.e.,
$\widetilde{S}_{min}\subset{H_a}\subset\widetilde{S}_{max}$, where
$\widetilde{S}_{max}=\widetilde{S}_{min}^\dag$ is the adjoint of
$\widetilde{S}_{min}$, see section \ref{subsec4.1}.

We show that spectral properties of $H_a$ are completely
characterized by the pair $\{a, \widetilde{W}_\lambda\}$,  where
$a\in\mathbb{C}$ distinguishes $H_a$ among all proper extensions of
$\widetilde{S}_{min}$, 
while the Weyl-Titchmarsh function $\widetilde{W}_\lambda$
(\ref{ggg14}) characterizes the symmetric operator
$\widetilde{S}_{min}$ which is `the common part' of all $H_a$; see
Theorems \ref{neww71}, \ref{ggg21}, and \ref{ggg16}.

One of interesting features of the model is fact that
$a\in\mathbb{C}$ determines the measure of non-self-adjointness of
the  operators $H_a$, while the choice of $q$ defines
the symmetric operator $\widetilde{S}_{min}$ and, therefore, the
structure of the holomorphic function $\widetilde{W}_\lambda$. Such
`a separation of responsibility' of parameters of the model allows
one to preserve its solvability and illustrate the possible
appearance of exceptional points and eigenvalues on continuous
spectrum, see Example \ref{ggg57} and subsec. \ref{subsec4.3}.

The proposed approach to the construction of non-self-adjoint
nonlocal point interaction models is not restricted to the case of
$\delta$-interactions only and it can be applied to the wider class
of ordinary point interaction models. We illustrate this point in
sections \ref{sec2} -- \ref{sec3} which are devoted to general
case of one point interactions including combinations of $\delta$- and
$\delta'$-interactions.

Throughout the paper, $\mathcal{D}(H)$, $\mathcal{R}(H)$, and $\ker{H}$ denote the
domain, the range, and the null-space of a linear operator $H$, respectively,
while $H\upharpoonright_{\mathcal{D}}$ stands for the restriction of
$H$ to the set $\mathcal{D}$. The adjoint of $H$ with respect to the
natural inner product $(\cdot,\cdot)$ (linear in the second
argument) in $L_2(\mathbb{R})$ is denoted by  $H^{\dag}$.

\section{One point interactions}\label{sec2}
\subsection{Ordinary one point interactions}\label{s1}

A one-dimensional Schr\"{o}dinger operator with interactions
supported at the point $x=0$ can be defined by the formal expression
 \begin{equation}\label{lesia11}
 -\frac{d^2}{dx^2}+a<\delta,\cdot>\delta(x)+b<\delta',\cdot>\delta(x)+
 c<\delta,\cdot>\delta'(x)+d<\delta',\cdot>\delta'(x),
 \end{equation}
where $\delta$ and $\delta'$ are, respectively, the Dirac $\delta$-function
and its derivative, the parameters $a,b,c,d$ are complex numbers, and
$$
<\delta,f>=f(0), \qquad <\delta',f>:=-f'(0), \qquad \forall{f}\in{W}_2^2(\mathbb{R}).
$$

Denote $\mathbf{T}=\left[\begin{array}{cc}
 a & b \\
 c & d
 \end{array}\right]$.  Then (\ref{lesia11}) can be rewritten in more compact form
\begin{equation}\label{bebe8}
-\frac{d^2}{dx^2}+[\delta, \delta']{\mathbf T}\left[\begin{array}{c}
 <\delta, \cdot> \\
 <\delta', \cdot>
 \end{array}\right].
\end{equation}

The expression (\ref{bebe8}) determines the symmetric (non-self-adjoint) operator
$$
S=-\frac{d^2}{dx^2}, \qquad  \mathcal{D}(S)=\{f\in{W}_2^2(\mathbb{R}) \ : \ f(0)=f'(0)=0\},
$$
in $L_2(\mathbb{R})$, which does not depend on the choice of $a,b,c,d$.
In order to take into account the impact of these parameters,
we should extend the action of $\delta$ and $\delta'$ onto ${W}_2^2(\mathbb{R}\backslash\{0\})$.
The most natural way is
$$
<\delta,f>:=f_r(0)=\displaystyle{\frac{f(0+)+f(0-)}{2}}, \quad
<\delta',f>:=-f_r'(0)=-\displaystyle{\frac{f'(0+)+f'(0-)}{2}}.
$$

Furthermore, we assume that the second derivative in (\ref{bebe8})
acts on ${W}_2^2(\mathbb{R}\backslash\{0\})$ in the distributional
sense, that is
$$
-{f''}=-\{f''(x)\}_{x\not=0}-f_s(0)\delta'(x)-f_s'(0)\delta(x),
\qquad f\in{W}_2^2(\mathbb{R}\backslash\{0\}),
$$
where
$$
f_s(0)=f(0+)-f(0-), \qquad f_s'(0)=f'(0+)-f'(0-).
$$
Then, the action of (\ref{bebe8}) on functions
$f\in{W}_2^2(\mathbb{R}\backslash\{0\})$ can be represented as
follows:
\begin{equation}\label{lesia12}
 -\{f''(x)\}_{x\not=0}+[\delta, \delta'][\mathbf{T}\Gamma_0f-\Gamma_1f],
 \end{equation}
 where
$$
\Gamma_0f=\left[\begin{array}{c}
 <\delta, f> \\
 <\delta', f>
 \end{array}\right]=\left[\begin{array}{c}
 f_r(0) \vspace{3mm} \\
-f_r'(0)
\end{array}\right], \qquad \Gamma_1f=\left[\begin{array}{c}
f_s'(0) \vspace{3mm} \\
f_s(0)
\end{array}\right].
$$
Obviously, (\ref{lesia12}) determines a function from $L_2(\mathbb{R})$
if and only if $\mathbf{T}\Gamma_0f=\Gamma_1f$.
Therefore, the expression (\ref{lesia11}) gives rise to the operator
$-{d^2}/{dx^2}$ in $L_2(\mathbb{R})$ with the domain of definition
$\{f\in{W}_2^2(\mathbb{R}\backslash\{0\}) \ : \ \mathbf{T}\Gamma_0f-\Gamma_1f=0\}$.

\subsection{Nonlocal one point interactions}
Let us generalize the one point interactions potential
considered in (\ref{lesia11}) by adding a nonlocal point interactions part
$$
<\delta,\cdot>q_1(x)+(q_1, \cdot)\delta(x)+
 (q_2, \cdot)\delta'(x)+<\delta',\cdot>q_2(x),
$$
where functions $q_j\in{L_2(\mathbb{R})}$ are assumed to be
piecewise continuous and $(\cdot,\cdot)$ is the standard inner
product (linear in the second argument) of ${L_2(\mathbb{R})}$. Then
the generalization of (\ref{bebe8}) takes the form
\begin{equation}\label{lesia11b}
-\frac{d^2}{dx^2}+[\delta, \delta']\left({\mathbf T}\left[\begin{array}{c}
 <\delta, \cdot> \\
 <\delta', \cdot>
 \end{array}\right]+\left[\begin{array}{c}
 (q_1, \cdot) \\
 (q_2, \cdot)
 \end{array}\right]\right)+[q_1, q_2]\left[\begin{array}{c}
 <\delta, \cdot> \\
 <\delta', \cdot>
 \end{array}\right].
 \end{equation}
Extending, by analogy with  (\ref{bebe8}), the action of
(\ref{lesia11b}) onto ${W}_2^2(\mathbb{R}\backslash\{0\})$  we
obtain
\begin{equation}\label{lesia12b}
-\{f''(x)\}_{x\not=0}+[\delta,
\delta'][\mathbf{T}\Gamma_0f-{\Gamma}_1f]+[q_1, q_2]\Gamma_0f,
\end{equation}
where
\begin{equation}\label{e2}
 \Gamma_0f=\left[\begin{array}{c}
 <\delta, f> \\
 <\delta', f>
 \end{array}\right]=\left[\begin{array}{c}
 f_r(0) \vspace{3mm} \\
-f_r'(0)
\end{array}\right], \quad \Gamma_1f=\left[\begin{array}{c}
f_s'(0)-(q_1, f) \vspace{3mm} \\
f_s(0)-(q_2, f)
\end{array}\right].
\end{equation}

The expression (\ref{lesia12b}) has sense as a function from $L_2(\mathbb{R})$ if
and only if the second term of (\ref{lesia12b}) is vanished, i.e.,
if \ $\mathbf{T}\Gamma_0f-{\Gamma}_1f=0.$ This means that the
formula (\ref{lesia11b}) determines the following operator in
$L_2(\mathbb{R})$:
\begin{equation}\label{new121}
H_{\mathbf{T}}f=-\frac{d^2f}{dx^2}+[q_1,q_2]\Gamma_0f=-\{f''(x)\}_{x\not=0}+f_r(0)q_1(x)-f_r'(0)q_2(x)
\end{equation}
with the domain of definition
\begin{equation}\label{e1b}
 \mathcal{D}(H_{\mathbf{T}})=\{f\in{W}_2^2(\mathbb{R}\backslash\{0\}) \ : \
 (\mathbf{T}\Gamma_0-\Gamma_1)f=0\}.
\end{equation}

The  maximal operator in the Hilbert space $L_2(\mathbb{R})$ that can be determined by
(\ref{lesia11b}) coincides with
\begin{equation}\label{e4}
S_{max}f=-\frac{d^2f}{dx^2}+[q_1,q_2]\Gamma_0f, \qquad
{f}\in\mathcal{D}(S_{max})={W}_2^2(\mathbb{R}\backslash\{0\}).
\end{equation}
Taking (\ref{e2}) into account, we obtain
$$
S_{max}f=-\{f''(x)\}_{x\not=0}+{f_r(0)}q_1-f_r'(0)q_2.
$$

The operator $S_{max}$ satisfies the Green's identity
\begin{equation}\label{bebe9}
(S_{max}f,g)-(f, S_{max}g)=(\Gamma_1f)\cdot\Gamma_0g-(\Gamma_0f)\cdot\Gamma_1g,
\end{equation}
where the dot $``\cdot"$ in the right hand side means the standard
inner product in $\mathbb{C}^2$. Moreover, according to \cite[Lemma
1]{Nizhnik}, for any vectors $h_0, h_1\in\mathbb{C}^2$, there
exists $f\in\mathcal{D}(S_{max})$  such that $\Gamma_0f=h_0$ and
$\Gamma_1f=h_1$.

The next operator plays an important role in what follows:
\begin{equation}\label{bebe5}
H_\infty=S_{max}\upharpoonright_{\mathcal{D}(H_{\infty})}, \qquad
\mathcal{D}(H_{\infty})=\{f\in\mathcal{D}(S_{max}) \ : \
\Gamma_0f=0 \}.
\end{equation}
In view of (\ref{e2}) and (\ref{e4}),
$$
H_\infty{f}=-\frac{d^2f}{dx^2}, \quad
f\in\mathcal{D}(H_\infty)=\{f\in{W}_2^2(\mathbb{R}\backslash\{0\}) \
: \ f_r(0)=f'_{r}(0)=0\}.
$$
It is easy to check that $H_\infty$ is a positive\footnote{since
$(H_\infty{f}, f)=\int_{\mathbb{R}}|f'(x)|^2dx>0$ for nonzero
$f\in\mathcal{D}(H_\infty)$} self-adjoint operator in
$L_2(\mathbb{R})$.

Due to \cite[Corollary 2.5]{Bernhdt}, the self-adjointness of
$H_\infty$, the Green identity (\ref{bebe9}), and the surjectivity
of the mapping $(\Gamma_0,\Gamma_1) :
\mathcal{D}(S_{max})\to\mathbb{C}^2\oplus\mathbb{C}^2$ lead to the
conclusion that the operator
$S_{min}=S_{max}\upharpoonright_{\mathcal{D}(S_{min})}$ with the domain
of definition $\mathcal{D}(S_{min})=\{f\in\mathcal{D}(S_{max}) \ : \
\Gamma_0f=\Gamma_1f=0\}$ is a closed symmetric operator in
$L_2(\mathbb{R})$. Precisely, $S_{min}{f}=-\frac{d^2f}{dx^2}$ with
the domain
\begin{equation}\label{ggg2}
\mathcal{D}(S_{min})=\left\{f\in{W}_2^2(\mathbb{R}\backslash\{0\})  :  \begin{array}{cc}
f_r(0)=0 &  f_s(0)=(q_2, f) \\
f_r'(0)=0 & f_s'(0)=(q_1, f)
\end{array}\right\}.
\end{equation}
Moreover, the relation $S_{min}^\dag=S_{max}$ holds and the
collection $(\mathbb{C}^2, \Gamma_0, \Gamma_1)$ is a boundary
triplet\footnote{see the Appendix} of $S_{max}$. The latter property
is especially important because operators $H_{\mathbf{T}}$, are
intermediate extensions between $S_{min}$ and $S_{max}$ and their
domains of definition are determined in terms of boundary operators
$\Gamma_j$, see (\ref{e1b}). Therefore, the well developed methods
of boundary triplet theory \cite{Schm}
 can be applied for the investigation of $H_{\mathbf{T}}$.

\section{Special cases of nonlocal one point interactions}
\subsection{Self-adjoint nonlocal one point interactions}

\begin{lemma}\label{au23}
If the entries of ${\mathbf{T}}$ satisfy the conditions $a, d
\in{\mathbb{R}}, \ b=c^*,$ then the corresponding operator
$H_{\mathbf{T}}$ defined by (\ref{new121}) is self-adjoint in
$L_2(\mathbb{R})$ for any choice of $q_j\in{L_2(\mathbb{R})}$.
\end{lemma}

\emph{Proof.}  It follows from the theory of boundary triplets (see
the Appendix) that  $H_{\mathbf{T}}^\dag=H_{{\mathbf{T}}^\dag}$,
where ${\mathbf{T}}^\dag={({\mathbf{T}}^*)^t}$. Therefore,
$H_{\mathbf{T}}$ is a self-adjoint operator if and only if the
matrix ${\mathbf{T}}$ is Hermitian. The latter is equivalent to the
conditions $a, d \in{\mathbb{R}}, \ b=c^*$. \rule{2mm}{2mm}

\subsection{{$\mathcal{PT}$}-symmetric nonlocal one point
interactions} As usual \cite{Carl} we consider the space parity
operator $\mathcal{P}f(x)=f(-x)$ and the conjugation operator
$\mathcal{T}f={f}^*$. An operator $H$ acting in $L_2(\mathbb{R})$ is
called ${\mathcal{PT}}$-symmetric if
${\mathcal{PT}}H=H{\mathcal{PT}}$.

\begin{lemma}\label{au3}
If the entries of ${\mathbf{T}}$ and the functions $q_j$ satisfy the conditions
\begin{equation}\label{new122}
a,d \in{\mathbb{R}}, \qquad b,c \in{i\mathbb{R}}, \qquad
\mathcal{PT}q_1=q_1, \qquad \mathcal{PT}q_2=-q_2,
\end{equation}
then the corresponding operator $H_{\mathbf{T}}$ defined by
(\ref{new121}) is $\mathcal{PT}$-symmetric.
\end{lemma}

\emph{Proof.} It is easy to check that, for any $f\in{W}_2^2(\mathbb{R}\backslash\{0\})$,
$$
({\mathcal{P}f})_r(0)=f_r(0), \quad ({\mathcal{P}f})_s(0)=-f_s(0),
\quad {(\mathcal{P}f)'}_r(0)=-f_r'(0), \quad
{(\mathcal{P}f)'}_s(0)=f_s'(0).
$$
These relations, the definition (\ref{e2}) of $\Gamma_j$, and (\ref{new122}) lead to the conclusion
that\footnote{The same symbol ${\mathcal{T}}$ are used for the
conjugation operators in $L_2(\mathbb{R})$ and in $\mathbb{C}^2$.}
\begin{equation}\label{bebe4}
\Gamma_j\mathcal{PT}f=\sigma_3{\mathcal{T}}\Gamma_jf, \quad \sigma_3=\left[\begin{array}{cc}
1 & 0 \\
0 & -1
\end{array} \right], \quad j=0,1.
\end{equation}
Therefore, if (\ref{new122}) holds, then the operator $S_{max}$ defined by (\ref{e4})
is $\mathcal{PT}$-symmetric
$$
\mathcal{PT}S_{max}f
=-\frac{d^2}{dx^2}\mathcal{PT}f+[q_1,q_2]\sigma_3{\mathcal{T}}\Gamma_0f=S_{max}\mathcal{PT}f.
$$

Since $H_{\mathbf{T}}$ is the restriction of $S_{max}$ onto
$\mathcal{D}(H_{\mathbf{T}})$, the invariance of
$\mathcal{D}(H_{\mathbf{T}})$ with respect to $\mathcal{PT}$ will
guarantee the $\mathcal{PT}$-symmetricity of $H_{\mathbf{T}}$.

Let us prove that $\mathcal{PT} : \mathcal{D}(H_{\mathbf{T}}) \to
\mathcal{D}(H_{\mathbf{T}})$.  To do that, we consider an arbitrary
$f\in\mathcal{D}(H_{\mathbf{T}})$. Then, according to (\ref{e1b}),
$\mathbf{T}\Gamma_0f=\Gamma_1f$ and the inclusion
$\mathcal{PT}f\in\mathcal{D}(H_{\mathbf{T}})$ is equivalent to the
condition $\mathbf{T}\Gamma_0\mathcal{PT}f=\Gamma_1\mathcal{PT}f$.
By virtue of (\ref{bebe4}),
$\mathbf{T}\Gamma_0\mathcal{PT}f=\mathbf{T}\sigma_3{\mathcal{T}}\Gamma_0f$
and $$
\Gamma_1\mathcal{PT}f=\sigma_3{\mathcal{T}}\Gamma_1f
=\sigma_3{\mathcal{T}}\mathbf{T}\Gamma_0f=\sigma_3{{\mathbf{T}}^*}{\mathcal{T}}\Gamma_0f.
$$
This means that the required identity
$\mathbf{T}\Gamma_0\mathcal{PT}f=\Gamma_1\mathcal{PT}f$ is true
if and only if $\mathbf{T}\sigma_3=\sigma_3{\mathbf{T}}^*$. The
latter matrix relation holds if the entries of $\mathbf{T}$ satisfy
(\ref{new122}). \rule{2mm}{2mm}

\subsection{{$\mathcal{P}$}-self-adjoint nonlocal one point
interactions} An operator $H_{\mathbf{T}}$ defined by
(\ref{new121}) is called $\mathcal{P}$-self-adjoint if
$\mathcal{P}H_{\mathbf{T}}=H_{\mathbf{T}}^\dag\mathcal{P}$.

\begin{lemma}\label{au3b}
If the entries of ${\mathbf{T}}$ and the functions $q_j$ satisfy the conditions
\begin{equation}\label{new124}
a,d \in{\mathbb{R}}, \qquad b=-{c}^* , \qquad \mathcal{P}q_1=q_1,
\qquad \mathcal{P}q_2=-q_2,
\end{equation}
then the operator $H_{\mathbf{T}}$ is {$\mathcal{P}$}-self-adjoint.
\end{lemma}

\emph{Proof.} Similarly to the proof of Lemma \ref{au3} we
check that $\Gamma_j\mathcal{P}f=\sigma_3\Gamma_jf$ and
show that the conditions (\ref{new124})
ensure the commutation relation  $S_{\max}\mathcal{P}=\mathcal{P}S_{max}$.

The operators $H_{\mathbf{T}}$ and $H_{\mathbf{T}}^\dag$ are
restrictions of $S_{max}$. Therefore, the condition $\mathcal{P} :
\mathcal{D}(H_{\mathbf{T}}) \to \mathcal{D}(H_{\mathbf{T}}^\dag)$
means the identity
$\mathcal{P}H_{\mathbf{T}}=H_{\mathbf{T}}^\dag\mathcal{P}$.

Let us verify that $\mathcal{P} : \mathcal{D}(H_{\mathbf{T}}) \to
\mathcal{D}(H_{\mathbf{T}}^\dag)$. Since
$H_{\mathbf{T}}^\dag=H_{{{\mathbf{T}}^*}^t}$, the domains of
definition $\mathcal{D}(H_{\mathbf{T}})$ and
$\mathcal{D}(H_{\mathbf{T}}^\dag)$  are determined by (\ref{e1b})
with the matrices ${\mathbf{T}}$ and ${{\mathbf{T}}^*}^t$,
respectively. Let  $f\in\mathcal{D}(H_{\mathbf{T}})$. Then
$\mathbf{T}\Gamma_0f=\Gamma_1f$ and the inclusion
$\mathcal{P}f\in\mathcal{D}(H_{\mathbf{T}}^\dag)$ is equivalent to
the condition
${\mathbf{T}^*}^t\Gamma_0\mathcal{P}f=\Gamma_1\mathcal{P}f$.

Taking into account  that $\Gamma_j\mathcal{P}f=\sigma_3\Gamma_jf$,
we obtain
 ${\mathbf{T}^*}^t\Gamma_0\mathcal{P}f={\mathbf{T}^*}^t\sigma_3\Gamma_0f$ and
 $\Gamma_1\mathcal{P}f=\sigma_3\Gamma_1f=\sigma_3\mathbf{T}\Gamma_0f$.
Hence, ${\mathbf{T}^*}^t\Gamma_0\mathcal{P}f=\Gamma_1\mathcal{P}f$
holds if and only if ${\mathbf{T}^*}^t\sigma_3=\sigma_3\mathbf{T}$.
This matrix relation holds if the entries $a,b,c,d$ of $\mathbf{T}$
satisfy (\ref{new124}). \rule{2mm}{2mm}

\section{Spectral Analysis of $H_{\mathbf{T}}$}\label{sec3}

The relations  (\ref{new121}), (\ref{e1b}) lead to the conclusion
that operators $H_{\mathbf{T}}$ are finite rank perturbations of the
self-adjoint operator $H_\infty$ defined by (\ref{bebe5}). The
spectrum of $H_\infty$ is purely continuous and it coincides with
$[0, \infty)$. This means that the continuous spectrum of each
$H_{\mathbf{T}}$ coincides with $[0, \infty)$ and only eigenvalues
of $H_{\mathbf{T}}$ may appear in $\mathbb{C}\setminus[0,\infty)$.

An eigenfunction of $H_{\mathbf{T}}$ should be the eigenfunction of $S_{max}$
corresponding to the same eigenvalue (since $S_{max}$ is an extension of
$H_{\mathbf{T}}$).

The kernel subspace $\ker(S_{max}-\lambda{I})$ has the dimension $2$
for any choice of $\lambda\in\mathbb{C}\setminus[0,\infty)$. Let
$u_\lambda, v_\lambda$ be a basis of $\ker(S_{max}-\lambda{I})$.
Then, any $f\in\ker(S_{max}-\lambda{I})$ has the form
$f=c_1u_\lambda+c_2v_\lambda$ and $f$ turns out to be the
eigenfunction of $H_{\mathbf{T}}$ corresponding to the eigenvalue
$\lambda$ if and only if $f$ belongs to the domain
$\mathcal{D}(H_{\mathbf{T}})$ determined by (\ref{e1b}), i.e., if
$c_1, c_2$ are nonzero solutions of the linear system
$$
c_1(\mathbf{T}\Gamma_0-\Gamma_1)u_\lambda+c_2(\mathbf{T}\Gamma_0-\Gamma_1)v_\lambda=0.
$$
Therefore, the eigenvalues $\lambda\in\mathbb{C}\setminus[0,\infty)$
of $H_{\mathbf{T}}$ coincide with the roots of the characteristic
equation
\begin{equation}\label{bebe6}
\det[(\mathbf{T}\Gamma_0-\Gamma_1)u_\lambda, (\mathbf{T}\Gamma_0-\Gamma_1)v_\lambda]=0.
\end{equation}

Let us assume, without loss of generality, that the eigenfunctions
$u_\lambda, v_\lambda$ are chosen in such a way that
$\Gamma_0{u_\lambda}=\left[\begin{array}{c}
1 \\
0 \end{array}\right]$ and $\Gamma_0{v_\lambda}=\left[\begin{array}{c}
0 \\
1 \end{array}\right]$. Then the characteristic equation
(\ref{bebe6}) for the determination of eigenvalues of
$H_{\mathbf{T}}$  takes the form
\begin{equation}\label{bebe7}
\det(\mathbf{T}-W_\lambda)=0,
\end{equation}
where $2\times{2}$-matrix $W_\lambda=[\Gamma_1u_\lambda,
\Gamma_1v_\lambda]$ is called \emph{the Weyl-Titchmarsh function
associated to the boundary triplet $(\mathbb{C}^2, \Gamma_0,
\Gamma_1)$}. The Weyl-Titchmarsh function $W_\lambda$ is holomorphic
on $\mathbb{C}\setminus[0,\infty)$ and it satisfies the relation
$(W_\lambda^*)^t=W_{\lambda^*}$ (see the Appendix).

\subsection{Eigenfunctions of $S_{max}$}
Let us write any $\lambda\in\mathbb{C}\setminus[0,\infty)$ as
$\lambda=k^2$, where $k\in\mathbb{C}_+=\{k\in\mathbb{C}  :  Im\ k>0 \}$ and
consider the function
$$
G(x)=\frac{i}{2k}e^{ik|x|}.
$$
Obviously, $G(\cdot)$ belongs to ${W}_2^2(\mathbb{R}\backslash\{0\})$ and
$$
-G''-k^2G=0, \qquad -(G')''-k^2G'=0, \qquad x\not=0.
$$
 Moreover,
$$
G_r(0)=\frac{i}{2k}, \quad G_r'(0)=0, \quad G_r''(0)=-\frac{ik}{2} \quad G_s(0)=0, \quad G_s'(0)=-1,  \quad G_s''(0)=0.
$$

The convolution
$$
f=(G\ast{q})(x)=\int_{-\infty}^{\infty} G(x-s)q(s) \ ds
$$
($q\in{L_2(\mathbb{R})}$ is a piecewise continuous function)
is the solution of the differential equation
$-f''-k^2f=q$ in $L_2(\mathbb{R})$.
\begin{lemma}\label{l1}
The functions
$$
\begin{array}{c}
u(x)=-(G\ast{q_1})(x)-2ik[1+(G\ast{q_1})(0)]G(x)+\frac{2i}{k}(G'\ast{q_1})(0)G'(x)
\vspace{3mm} \\
v(x)=-(G\ast{q_2})(x)-2ik(G\ast{q_2})(0)G(x)-\frac{2i}{k}[1-(G'\ast{q_2})(0)]G'(x)
\end{array}
$$
form the basis of the eigenfunction subspace $\ker(S_{max}-k^2I)$.
\end{lemma}
\emph{Proof.} An elementary analysis shows that the functions $u, v$
belong to ${W}_2^2(\mathbb{R}\backslash\{0\})$  and
\begin{equation}\label{au4}
\begin{array}{l}
u_r(0)=1, \quad u_s(0)=-\frac{2i}{k}(G'\ast{q_1})(0),  \quad v_r(0)=0, \quad v_s(0)
=\frac{2i}{k}[1-(G'\ast{q_2})(0)] \vspace{3mm} \\
u'_r(0)=0, \quad  u'_s(0)=2ik[1+(G\ast{q_1})(0)], \quad v'_r(0)=-1,
\quad v'_s(0)=2ik(G\ast{q_2})(0)
\end{array}
\end{equation}

The first and the third columns in (\ref{au4}) mean that $u$ and
$v$ are linearly independent and $\Gamma_0{u}=\left[\begin{array}{c}
1 \\
0 \end{array}\right],$ \ $\Gamma_0{v}=\left[\begin{array}{c}
0 \\
1 \end{array}\right]$.
 Furthermore, taking into account (\ref{e4}) and (\ref{au4}) we obtain
for almost all $x\in\mathbb{R}$
$$
(S_{max}-k^2I)u=-u''-k^2u+q_1=-q_1+q_1=0.
$$
Similarly, $(S_{max}-k^2I)v=-v''-k^2v+q_2=-q_2+q_2=0.$
Hence, the functions $u, v$ belong to $\ker(S_{max}-k^2I)$ and
they form a basis of this subspace.
\rule{2mm}{2mm}

\subsection{The Weyl-Titchmarsh function associated to $(\mathbb{C}^2, \Gamma_0,
\Gamma_1)$}
Since $\Gamma_0{u}=\left[\begin{array}{c}
1 \\
0 \end{array}\right]$ and $\Gamma_0{v}=\left[\begin{array}{c}
0 \\
1 \end{array}\right]$, the  Weyl-Titchmarsh function associated to $(\mathbb{C}^2, \Gamma_0,
\Gamma_1)$ has the form $W_\lambda=[\Gamma_1u, \Gamma_1v]$, where,
in view of (\ref{e2}) and (\ref{au4}),
$$
\Gamma_1u=\left[\begin{array}{c}
2ik[1+(G\ast{q_1})(0)]-(q_1, u) \\
-\frac{2i}{k}(G'\ast{q_1})(0)-(q_2, u) \end{array}\right],
\
\Gamma_1v=\left[\begin{array}{c}
2ik(G\ast{q_2})(0)-(q_1,v) \\
\frac{2i}{k}[1-(G'\ast{q_2})(0)]-(q_2,v) \end{array}\right].
$$
 Making some additional rudimentary calculations (mainly related to the
calculation of scalar products $(q, u)$, $(q, v)$ for functions $u,
v$ from Lemma \ref{l1}), we obtain
\begin{equation}\label{bbb5}
W_\lambda=\left[\begin{array}{cc}
({q}_1, G\ast{q_1}) & ({q}_1, G\ast{q_2})   \\
 ({q}_2, G\ast{q_1}) & ({q}_2, G\ast{q_2})
\end{array}\right] + \left[\begin{array}{cc}
r_{11} & r_{12}  \\
 r_{21} & r_{22}
\end{array}\right],
\end{equation}
where
$$
r_{11}=2ik[1+(G\ast{q_1})(0)][1+(G\ast{{q}_1^*})(0)]+\frac{2i}{k}(G'\ast{q_1})(0)(G'\ast{{q}_1^*})(0),
$$
$$
 r_{22}=\frac{2i}{k}[1-(G'\ast{q_2})(0)][1-(G'\ast{{q}_2^*})(0)]+2ik(G\ast{q_2})(0)(G\ast{{q}_2^*})(0),
$$
$$
r_{12}=2ik(G\ast{q_2})(0)[1
+(G\ast{{q}_1^*})(0)]-\frac{2i}{k}(G'\ast{{q}_1^*})(0)[1-(G'\ast{{q}_2})(0)],
$$
$$
r_{21}=2ik(G\ast{{q}_2^*})(0)[1
+(G\ast{{q}_1})(0)]-\frac{2i}{k}(G'\ast{{q}_1})(0)[1-(G'\ast{{q}_2^*})(0)].
$$
Denote
$$
B_{q_1,q_2}=\left[\begin{array}{cc}
1+(G\ast{{q}_1})(0) & (G\ast{{q}_2})(0) \\
-(G'\ast{{q}_1})(0) &  1-(G'\ast{{q}_2})(0)
\end{array}\right].
$$
Then (\ref{bbb5}) can be rewritten as follows:
\begin{equation}\label{bbb5c}
W_\lambda=\left[\begin{array}{cc}
({q}_1, G\ast{q_1}) & ({q}_1, G\ast{q_2})   \\
 ({q}_2, G\ast{q_1}) & ({q}_2, G\ast{q_2})
\end{array}\right]
 + B_{q_1^*,q_2^*}^t\left[\begin{array}{cc}
2ik & 0 \vspace{3mm}  \\
 0 & \displaystyle{\frac{2i}{k}}
\end{array}\right]B_{q_1,q_2}.
\end{equation}

Substituting (\ref{bbb5c}) into (\ref{bebe7}) we obtain the
characteristic equation for eigenvalues
$\lambda\in\mathbb{C}\setminus[0,\infty)$ of $H_{\mathbf{T}}$. In
particular, if $q_1=q_2=0$, the Weyl function $W_\lambda$ coincides
with $\left[\begin{array}{cc}
2ik & 0   \\
 0 & {{2i}/{k}}
\end{array}\right]$ and the equation (\ref{bebe7}) is transformed to the polynomial
\begin{equation}\label{ggg4}
2dk^2+ik(\det{\mathbf{T}}-4)+2a=0,
\end{equation}
which determines spectra of ordinary point interactions considered
in subsection \ref{s1}.

\section{Nonlocal $\delta$-interaction}\label{sec4}

\subsection{Definition and description of eigenvalues}\label{subsec4.1}

The classical one point $\delta$-interaction is given by the formal expression
\begin{equation}\label{au2}
-\frac{d^2}{dx^2}+a<\delta,\cdot>\delta(x), \qquad a\in\mathbb{C}
\end{equation}

It is natural to suppose that the generalization of (\ref{au2}) to
the \emph{nonlocal} case consists in the addition of the nonlocal
part $<\delta,\cdot>q(x)+(q, \cdot)\delta(x)$ of
$\delta$-interaction. For this reason, a nonlocal one-point
$\delta$-interaction can be defined via the formal expression
$$
-\frac{d^2}{dx^2}+a<\delta,\cdot>\delta(x)+<\delta,\cdot>q(x)+(q,\cdot)\delta(x),
 \quad a\in\mathbb{C}, \ q\in{L_2(\mathbb{R})},
$$
which is a particular case of (\ref{lesia11b}) with $\mathbf{T}=\left[\begin{array}{cc}
 a & 0 \\
 0 & 0
\end{array}\right]$, $q_1=q$, and $q_2=0$.
This means that the corresponding operator $H_{\mathbf{T}}\equiv{H}_{a}$ defined by
(\ref{new121}) and (\ref{e1b}) acts as
\begin{equation}\label{ggg}
H_{a}f=-\frac{d^2f}{dx^2}+f_r(0)q(x),
\end{equation}
on the domain of definition
\begin{equation}\label{ggg1}
 \mathcal{D}(H_{a})=\left\{f\in{W}_2^2(\mathbb{R}\backslash\{0\}) \ : \
 \begin{array}{l}
 f_s(0)=0  \vspace{2mm} \\
 f_s'(0)=af_r(0)+(q, f) \end{array} \right\}
\end{equation}

In view of Lemma \ref{au3}, the operator $H_{a}$ is
${\mathcal{PT}}$-symmetric if $a\in\mathbb{R}$ and
${\mathcal{PT}}q=q$. In this case, due to Lemma \ref{au23}, the
operator $H_{a}$ should be self-adjoint. Therefore,
$\mathcal{PT}$-symmetric nonlocal $\delta$-interactions are realized
via self-adjoint operators. The same result is true for the case of
$\mathcal{P}$-self-adjoint operators $H_{a}$ (see Lemma \ref{au3b}).

\begin{theorem}\label{neww71}
The operator $H_{a}$ defined by (\ref{ggg}) has an eigenvalue
$\lambda=k^2\in\mathbb{C}\setminus[0, \infty)$  if and only the following relation holds:
\begin{equation}\label{new134}
a=({q}, G\ast{q})+2ik[1+(G\ast{q})(0)][1+(G\ast{{q}^*})(0)], \quad k\in\mathbb{C}_+.
\end{equation}
\end{theorem}
\emph{Proof.} If $q=q_1$ and $q_2=0$, then the Weyl-Titchmarsh
function (\ref{bbb5c}) has the form
$$
W_\lambda=\left[\begin{array}{cc} ({q}, G\ast{q})+r_{11} &
-\frac{2i}{k}(G'\ast{{q}^*})(0) \vspace{3mm}  \\
 -\frac{2i}{k}(G'\ast{{q}})(0) & \frac{2i}{k}
\end{array}\right],
$$
where
$r_{11}=2ik[1+(G\ast{q})(0)][1+(G\ast{{q}^*})(0)]
+\frac{2i}{k}(G'\ast{q})(0)(G'\ast{{q}^*})(0).$

By virtue of (\ref{bebe7}), $\lambda\in\sigma_p(H_a)$ if and only if
$\det(\mathbf{T}-W_\lambda)=0$, where
$\mathbf{T}=\left[\begin{array}{cc}
 a & 0 \\
 0 & 0
\end{array}\right]$.  The direct calculation of $\det(\mathbf{T}-W_\lambda)$ in the latter equation
gives
(\ref{new134}).
\rule{2mm}{2mm}

Each operator $H_a$ satisfies the relation
$S_{min}\subset{H_a}\subset{S_{max}}$ because $H_a=H_{\mathbf{T}}$
with the matrix $\mathbf{T}$ determined above. This important
general relation (which holds for any $H_{\mathbf{T}}$) can be made
more precise for the particular case of operators $H_a$. Indeed, it
follows from (\ref{ggg1}) that $H_a$ are extensions of
the following operator:
\begin{equation}\label{new131}
\widetilde{S}_{min}{f}
=-\frac{d^2f}{dx^2}, \quad \mathcal{D}(\widetilde{S}_{min})
=\left\{f\in{W}_2^2(\mathbb{R}\backslash\{0\}) :  \begin{array}{c}
f_s(0)=f_r(0)=0 \\
f_s'(0)=(q, f)
\end{array}\right\}.
\end{equation}

It is easy to see (comparing $\mathcal{D}(\widetilde{S}_{min})$ with
the domain $\mathcal{D}({S}_{min})$ determined by (\ref{ggg2})) that
$\widetilde{S}_{min}$ is an extension of ${S}_{min}$, i.e.,
${S}_{min}\subset\widetilde{S}_{min}.$ Moreover, the operator
$\widetilde{S}_{min}$ is symmetric. This fact follows from the Green
identity (\ref{bebe7}) because $\Gamma_1f=0$ for all
$f\in\mathcal{D}(\widetilde{S}_{min})$.

Denote $\widetilde{S}_{max}=\widetilde{S}_{min}^\dag.$  The
calculation of the  adjoint operator gives
$$
\widetilde{S}_{max}f=-\frac{d^2f}{dx^2}+f_r(0)q(x),
 \quad  \mathcal{D}(\widetilde{S}_{max})=\{f\in{W}_2^2(\mathbb{R}\backslash\{0\}) \ : \
 f_s(0)=0\}.
$$

It is easy to check that
${S}_{min}\subset\widetilde{S}_{min}\subset{H_a}\subset\widetilde{S}_{max}\subset{S}_{max}.$
Thus, $H_a$ is a proper extension of the symmetric
operator $\widetilde{S}_{min}$. Furthermore, an elementary analysis shows
that:

$(i)$ the kernel subspace $\ker(\widetilde{S}_{max}-\lambda{I})$ is one-dimensional
and it is generated by the function (cf. Lemma \ref{l1})
\begin{equation}\label{ggg12}
u_{\lambda}(x)=-(G\ast{q})(x)-2ik[1+(G\ast{q})(0)]G(x);
\end{equation}

$(ii)$ the triple $(\mathbb{C}, \widetilde{\Gamma}_0, \widetilde{\Gamma}_1)$, where
\begin{equation}\label{au5}
\widetilde{\Gamma}_0f=f_r(0), \qquad \widetilde{\Gamma}_1f=f_s'(0)-(q,f),
 \qquad f\in\mathcal{D}(\widetilde{S}_{max})
\end{equation}
is the boundary triplet of $\widetilde{S}_{max}$ and
\begin{equation}\label{ggg13}
\widetilde{\Gamma}_0u_\lambda=1, \quad \widetilde{\Gamma}_1u_\lambda
=({q}, G\ast{q})+2ik[1+(G\ast{q})(0)][1+(G\ast{{q}^*})(0)],
\end{equation}
where $u_{\lambda}$ is determined by (\ref{ggg12});

$(iii)$ the operators $H_a$ initially defined by (\ref{ggg}) and
(\ref{ggg1}) can be rewritten in terms of the boundary triplet
$(\mathbb{C}, \widetilde{\Gamma}_0, \widetilde{\Gamma}_1)$ (cf.
(\ref{e1b})):
\begin{equation}\label{e1bc}
H_a=\widetilde{S}_{max}\upharpoonright\mathcal{D}(H_a),
 \quad \mathcal{D}(H_a)=\{f\in\mathcal{D}(\widetilde{S}_{max}) \ :
  \ (a\widetilde{\Gamma}_0-\widetilde{\Gamma}_1)f=0\};
\end{equation}

$(iv)$ the operator (cf. (\ref{bebe5}))
$$
\widetilde{H}_\infty=\widetilde{S}_{max}\upharpoonright\mathcal{D}(\widetilde{H}_\infty),
\qquad
\mathcal{D}(\widetilde{H}_\infty)=\{f\in\mathcal{D}(\widetilde{S}_{max})
\ : \   \widetilde{\Gamma}_0f=0 \}
$$
is positive self-adjoint and its spectrum coincides with $[0,\infty)$.

The items $(i)-(iv)$ allow one to simplify the investigation of
$H_a$. First of all we note that the Weyl-Titchmarsh function
$\widetilde{W}_\lambda$ associated to the boundary triplet
$(\mathbb{C}, \widetilde{\Gamma}_0, \widetilde{\Gamma}_1)$ is a
holomorphic function on
$\rho(\widetilde{H}_\infty)=\mathbb{C}\setminus[0,\infty)$ and, due
to (\ref{ggg13}), it has the form
\begin{equation}{\label{ggg14}}
\widetilde{W}_\lambda=\widetilde{\Gamma}_1u_\lambda=({q},
G\ast{q})+2ik[1+(G\ast{q})(0)][1+(G\ast{{q}^*})(0)].
\end{equation}

The obtained formula immediately justifies (\ref{new134}) because
$\lambda\in\mathbb{C}\setminus[0,\infty)$ is an eigenvalue of $H_a$
if and only if $\det(a-\widetilde{W}_\lambda)=0$ or, that is
equivalent, if $a=\widetilde{W}_\lambda$. The latter identity shows
that at least one of subspaces $\mathbb{C}_\pm$ belongs to
$\rho(H_a)$. Indeed, if $a\in\mathbb{R}$, then
$\rho(H_a)\supset\mathbb{C}_\pm$. If
$a\in\mathbb{C}\setminus\mathbb{R}$, then only non-real eigenvalues
of $H_a$ might be in $\mathbb{C}_\pm$. Let us assume that
$\lambda_{\pm}\in\sigma_p(H_a)$ with $Im\ \lambda_+>0$ and $Im\
\lambda_-<0$. Then, simultaneously, $Im\ a>0$ and $Im\ a<0$ (since
$\widetilde{W}_{\lambda\pm}=a$  and $(Im\ \lambda)(Im \
\widetilde{W}_\lambda)>0$ for $Im\ \lambda\not=0$, see the Appendix)
that is impossible. Therefore, at least one of $\mathbb{C}_\pm$ does
not belong to $\sigma(H_a)$. This result is not true for the general
case of one point interactions considered in section \ref{sec2}. For
instance, if $q_1=q_2=0$ and $a=d=0,$ \ $bc=4$, then the
characteristic equation (\ref{ggg4}) is vanished and the eigenvalues of
$H_{\mathbf{T}}$ fill the whole domain
$\mathbb{C}\setminus[0,\infty)$.

\begin{corollary}\label{ggg6}
The existence of a real eigenvalue of $H_{a}$ means that $H_{a}$ is
a self-adjoint operator in $L_2(\mathbb{R})$.
\end{corollary}
\emph{Proof.} Let $u_\lambda\in{L_2(\mathbb{R})}$ be an eigenfunction of
$H_a$ corresponding to a real eigenvalue $\lambda$. It follows
from the definition of $\widetilde{S}_{min}$ that
$\ker(\widetilde{S}_{min}-\lambda{I})=\{0\}$. Therefore, the
domain of $H_a$ can be represented as
$$
\mathcal{D}(H_a)=\{f=v+c{u_\lambda} \ : \ v\in\mathcal{D}(\widetilde{S}_{min}), \ c\in\mathbb{C} \}
$$
(since the symmetric operator $\widetilde{S}_{min}$ has the defect
index $1$) and
$$
H_af=H_a(v+c{u_\lambda})=\widetilde{S}_{min}v+\lambda{cu_\lambda}.
$$
Using the last expression we check that $Im \ (H_af,f)=0$ for all
$f=v+c{u_\lambda}$ from the domain of $H_a$. Therefore, $H_a$ is a
self-adjoint operator. \rule{2mm}{2mm}

\smallskip

In contrast to the case of ordinary one point interactions
considered in subsec. \ref{s1}, the operators $H_a$ may have real
eigenvalues embedded into continuous spectrum $[0,\infty)$. To see
this we rewrite the function $u_\lambda$ in (\ref{ggg12}) as
follows:
\begin{equation}\label{ggg41}
u_{\lambda}(x)=\left\{\begin{array}{l}
A_k(x)e^{ikx}+B_k(x)e^{-ikx}, \quad x>0 \\
C_k(x)e^{ikx}+D_k(x)e^{-ikx}, \quad x<0
\end{array}\right., \quad \lambda=k^2,
\end{equation}
where
$$
A_k(x)
=1+\frac{i}{2k}\int_0^\infty{e^{iks}}q(s)ds-\frac{i}{2k}\int_0^x{e^{-iks}}q(s)ds,
$$
$$
D_k(x)
=1+\frac{i}{2k}\int^0_{-\infty}{e^{-iks}}q(s)ds-\frac{i}{2k}\int^0_x{e^{iks}}q(s)ds,
$$
$$
B_k(x)=-\frac{i}{2k}\int_x^\infty{e^{iks}}q(s)ds, \quad
C_k(x)=-\frac{i}{2k}\int_{-\infty}^x{e^{-iks}}q(s)ds.
$$

If $\lambda=k^2$ with $k\in\mathbb{C_+}$, then the function
$u_\lambda$ belongs to $L_2(\mathbb{R})$ and it solves the
differential equation $-f''(x)+f_r(0)q(x)={\lambda}f(x)$ for
$x\not=0$. According to (\ref{ggg13}) and (\ref{ggg14}),
$u_\lambda$ belongs to the domain of definition (\ref{ggg1}) of the
operator $H_a$  with $a=\widetilde{W}_\lambda$. In other words,
$u_\lambda$  is the eigenfunction of $H_a$.

If $\lambda=k^2$ with $k\in\mathbb{R}\setminus\{0\}$, then the
function $u_\lambda$ defined by (\ref{ggg41}) turns out to be
\emph{generalized eigenfunction} of $H_a$. This means that
$u_{\lambda}$ preserves all properties above except the property of
being in $L_2(\mathbb{R})$. It should be noted that $u_\lambda$ may
belong to $L_2(\mathbb{R})$. In this case the generalized
eigenfunction coincides with the ordinary eigenfunction and  the
corresponding operator $H_a$ will have a positive eigenvalue
$\lambda=k^2$ located on continuous spectrum $[0,\infty)$. In view
of Corollary \ref{ggg6} this phenomenon is possible only for
self-adjoint operators $H_a$.

\begin{example}\label{ggg57} The case of an even function with finite support.
\\
Let $q$ be an even function with support in $[-\rho,\rho]$.
 The elementary calculation in (\ref{ggg41}) gives  that for all $|x|>\rho$
 $$
 u_\lambda(x)={\beta_k}e^{ik|x|}, \qquad  \beta_k=1-\frac{1}{k}\int_{0}^{\rho}\sin{ks}\
 q(s)ds.
 $$
It is easy to see that $u_\lambda$ will be in $L_2(\mathbb{R})$ if
and only if $\beta_k=0$. If $k\in\mathbb{R}\setminus\{0\}$ is a
solution of the last equation, then $u_\lambda$ turns out to be an
eigenfunction of the self-adjoint operator $H_a$, where
$a=\widetilde{W}_\lambda$ and $\widetilde{W}_\lambda$ is formally
defined by (\ref{ggg14}) with $\lambda=k^2\in(0,\infty)$.

It should be noted that the case of odd functions with finite
support is completely different. Indeed, if $q$ is odd with the
support in $[-\rho, \rho]$, then
$$
u_\lambda(x)=\left\{
\begin{array}{l}
(1-\frac{1}{k}\int_{0}^{\rho}\sin{ks}\ q(s)ds)e^{ikx}, \quad  x>\rho \\
(1+\frac{1}{k}\int_{0}^{\rho}\sin{ks}\ q(s)ds)e^{-ikx}, \quad  x<-\rho
\end{array}\right.
$$
Obviously, such a function $u_\lambda$ does not belong to
$L_2(\mathbb{R})$ and it cannot be an eigenfunction of $H_a$.
Therefore, in the case of odd function $q$ with finite support, the
corresponding operators  $H_a$ $(a\in\mathbb{C})$ have no positive
eigenvalues.

Let us consider the simplest example of even function
\begin{equation}\label{ggg58}
q(x)=Z\chi_{[-\rho,\rho]}(x)=\left\{
\begin{array}{l}
Z, \quad x\in[-\rho,\rho] \\
0, \quad x\in\mathbb{R}\setminus[-\rho,\rho]
\end{array}\right. \quad Z\in\mathbb{R}, \quad \rho>0.
\end{equation}
The characteristic equation $\beta_k=0$ takes the form $Z(1-\cos{k\rho})=k^2$.
Let $k_0\in\mathbb{R}\setminus\{0\}$ be the solution of this equation. Then
the function
$$
u_\lambda(x)=\frac{Z(1-\cos{k_0(\rho-|x|)})}{k_0^2}\chi_{[-\rho,\rho]}(x)
\qquad \lambda=k^2_0,
$$
belongs to the domain of definition
$$
\mathcal{D}(H_a)=\left\{f\in{W}_2^2(\mathbb{R}\backslash\{0\}) \ : \  \begin{array}{l}
 f(0-)=f(0+)\equiv{f(0)}  \vspace{2mm} \\
 f'(0+)-f'(0-)=af(0)+Z\int_{-\rho}^{\rho}f(x)dx \end{array} \right\}
$$
of the self-adjoint operator
$H_{a}f=-\frac{d^2f}{dx^2}+Zf(0)\chi_{[-\rho,\rho]}(x)$, where
$$
a=[u_\lambda']_s(0)-Z\int_{-\rho}^{\rho}u_{\lambda}(x)dx
=\frac{Z^2}{k_0^2}\left(\frac{\sin{2k_0\rho}}{k_0}-2\rho\right).
$$
The function $u_\lambda$ is an eigenfunction of $H_a$ corresponding
to the positive eigenvalue $\lambda=k^2_0$.
\end{example}

\subsection{Exceptional points}
The geometric multiplicity of any $\lambda\in\sigma_p(H_a)$ is $1$
due to $(i)$ and the fact that
$\ker(\widetilde{S}_{min}-\lambda{I})=\{0\}$. The algebraical
multiplicity can be calculated with the use of general formula
(\ref{ggg3}).

An eigenvalue of $H_a$ is called \emph{exceptional point} if its
geometrical multiplicity does not coincide with the algebraic
multiplicity. The presence of an exceptional point means that $H_a$
cannot be self-adjoint for any choice of inner product in
$L_2(\mathbb{R})$. By virtue of Corollary \ref{ggg6},  the
operators $H_a$ may only have non-real exceptional points.

\begin{theorem}\label{ggg21}
A non-real eigenvalue $\lambda_0$ of $H_a$ is an exceptional point
if and only if \ $\widetilde{W}_{\lambda_0}'=0$, where
$\widetilde{W}_{\lambda}'=\frac{d}{d\lambda}\widetilde{W}_{\lambda}$.
\end{theorem}

\emph{Proof.} The resolvent $(\widetilde{H}_\infty-{\lambda}I)^{-1}$
of a self-adjoint operator $\widetilde{H}_\infty$ is a holomorphic operator-valued function on
$\rho(\widetilde{H}_\infty)=\mathbb{C}\setminus[0,\infty)$. On the
other hand,  the resolvent $(H_a-\lambda{I})^{-1}$ may be a
meromorphic function on $\mathbb{C}\setminus[0,\infty)$ and its
poles are eigenvalues of $H_a$.

Let $\lambda_0\in\mathbb{C}\setminus\mathbb{R}$ be a pole of
$(H_a-\lambda{I})^{-1}$. Then its order coincides with the maximal
length of Jordan vectors associated with $\lambda_0$ (see, e.g.,
\cite[Chapt. 2]{WOL}). Therefore, the existence of an exceptional
point $\lambda_0$ of $H_a$ is equivalent to the existence of pole
$\lambda_0$ of order greater than one for the meromorphic
operator-valued function
\begin{equation}\label{bbb9}
\Xi(\lambda)=(H_a-\lambda{I})^{-1}-(\widetilde{H}_\infty-\lambda{I})^{-1}.
\end{equation}
In other words, $\lambda_0$ turns out to be an exceptional point of
$H_a$ if and only if there exists $v\in{L_2(\mathbb{R})}$ such that
\begin{equation}\label{ggg10}
\lim_{\lambda\to\lambda_0}\|(\lambda-\lambda_0)\Xi(\lambda)v\|=\infty.
\end{equation}

It is sufficient to suppose in (\ref{ggg10}) that
$v=u_{\lambda^*}\in\ker(\widetilde{S}_{max}-\lambda^*{I})$ (since
$H_a$ and $\widetilde{H}_\infty$ are extensions of
$\widetilde{S}_{min}$ and, hence,
$\Xi(\lambda)\upharpoonright_{\mathcal{R}(\widetilde{S}_{min}-\lambda{I})}=0$).

It follows from the Krein-Naimark resolvent formula (\ref{ggg9}) that
\begin{equation}\label{ggg17}
\|(\lambda-\lambda_0)\Xi(\lambda){u}_{\lambda^*}\|
=\left|\frac{\lambda-\lambda_0}{a-\widetilde{W}_{\lambda}}\right|\|
\gamma(\lambda)\gamma(\lambda^*)^\dag{u}_{\lambda^*}\|.
\end{equation}

Let us evaluate the part
$\|\gamma(\lambda)\gamma(\lambda^*)^\dag{u}_{\lambda^*}\|$ in
(\ref{ggg17}). In view of (\ref{ggg11}),
$$
\gamma(\lambda^*)^\dag{u}_{\lambda^*}
=\widetilde{\Gamma}_1(\widetilde{H}_{\infty}-\lambda{I})^{-1}{u}_{\lambda^*}.
$$
The operator $\widetilde{H}_{\infty}$ is defined in $(iv)$ and it
acts as $\widetilde{H}_{\infty}f=-\frac{d^2f}{dx^2}$ for all
functions
$f\in\mathcal{D}(\widetilde{H}_{\infty})=\{f\in{W}_2^2(\mathbb{R}\backslash\{0\})
\ : \ f(0-)=f(0+)=0\}.$ The resolvent of $\widetilde{H}_{\infty}$ is
well known and it takes especially simple form  for
$f={u}_{\lambda^*}$
$$
(\widetilde{H}_{\infty}-\lambda{I})^{-1}{u}_{\lambda^*}=\frac{1}{2i(Im
\ \lambda)}(u_\lambda-{u}_{\lambda^*}).
$$

The definition of the Weyl-Titchmarsh function
$\widetilde{W}_\lambda$ associated to the boundary triplet
$(\mathbb{C}, \widetilde{\Gamma}_0, \widetilde{\Gamma}_1)$ and the
relation $\widetilde{\Gamma}_0u_\lambda=1$ in (\ref{ggg13}) imply
that $\widetilde{\Gamma}_1u_\lambda=\widetilde{W}_\lambda$ for all
$\lambda\in\mathbb{C}\setminus[0,\infty)$. Therefore,
$$
\gamma(\lambda^*)^\dag{u}_{\lambda^*}
=\widetilde{\Gamma}_1(\widetilde{H}_{\infty}-\lambda{I})^{-1}{u}_{\lambda^*}
=\frac{\widetilde{\Gamma}_1(u_\lambda-{u}_{\lambda^*})}{2i(Im \ \lambda)}
=\frac{\widetilde{W}_\lambda-\widetilde{W}_{\lambda^*}}{2i(Im \ \lambda)}
=\frac{Im \ \widetilde{W}_\lambda}{Im \ \lambda}.
$$

Further, it follows from the definition of $\gamma$-field
$\gamma(\cdot)$ associated with $(\mathbb{C}, \widetilde{\Gamma}_0,
\widetilde{\Gamma}_1)$ (the Appendix) and (\ref{ggg13}) that
$\gamma(\lambda)c=cu_\lambda$ for all $c\in\mathbb{C}$. Hence,
$\gamma(\lambda)\gamma(\lambda^*)^\dag{u}_{\lambda^*}=\frac{Im \
\widetilde{W}_\lambda}{Im \ \lambda}u_\lambda$. Setting
$f_\lambda=u_\lambda$ in (\ref{ggg8}) we decide that
\begin{equation}\label{ggg22}
\|u_\lambda\|^2=\frac{Im \ \widetilde{W}_\lambda}{Im \ \lambda},
\qquad \lambda\in\mathbb{C}\setminus\mathbb{R}.
\end{equation}
Therefore,
$$
\alpha(\lambda)=\|\gamma(\lambda)\gamma(\lambda^*)^\dag{u}_{\lambda^*}\|
=\left(\frac{Im \ \widetilde{W}_\lambda}{Im \ \lambda}\right)^{3/2}.
$$

The function $\alpha(\lambda)$ is continuous in a neighborhood of
the non-real point $\lambda_0$  and $\alpha(\lambda_0)\not=0$.
Therefore, taking (\ref{ggg17}) into account, we decide that
(\ref{ggg10}) is equivalent to the condition
$$
\lim_{\lambda\to\lambda_0}\frac{a-\widetilde{W}_{\lambda}}{\lambda-\lambda_0}=0.
$$
Remembering that $a=\widetilde{W}_{\lambda_0}$ (since $\lambda_0$ is
an eigenvalue of $H_a$) we complete the proof. \rule{2mm}{2mm}
\begin{corollary}\label{au10}
If $H_a$ has an exceptional point $\lambda_0$, then $\lambda_0^*$ is
an exceptional point for $H_{a^*}$
\end{corollary}

The proof follows from Theorem \ref{ggg21} and the relation
$\widetilde{W}_\lambda^*=\widetilde{W}_{{\lambda^*}}$.

\smallskip

\subsection{Spectral singularities}
Let $H_a$ be \emph{a non-self-adjoint} operator with real spectrum.
The operator $H_a$ cannot have real eigenvalues due to Corollary
\ref{ggg6}. Therefore, the spectrum of $H_a$ is continuous and it
coincides with $[0,\infty)$.

If $H_a$ turns out to be self-adjoint with respect to an
appropriative choice of inner product of $L_2(\mathbb{R})$  (i.e, if
$H_a$ is similar to a self-adjoint operator in $L_2(\mathbb{R})$),
then its resolvent $(H_a-\lambda{I})^{-1}$  should satisfy the
standard evaluation
\begin{equation}\label{ggg15}
\|(H_a-\lambda{I})^{-1}f\|\leq\frac{C}{|Im\ \lambda|}\|f\|,
\end{equation}
where $C>0$ does not depend on
$\lambda\in\mathbb{C}\setminus\mathbb{R}$ and
$f\in{L_2}(\mathbb{R})$.

The case where $H_a$ is not similar to a self-adjoint operator in
$L_2(\mathbb{R})$ deals with the existence of special spectral
points of $H_a$ which are impossible for the spectra of self-adjoint
operators. Traditionally, these spectral points are called
\emph{spectral singularities} if they are located at the continuous
spectrum of $H_a$. Particular role pertaining to the spectral
singularities was discovered for the first time by Naimark
\cite{Naim}. Nowadays, various aspects of spectral singularities
including the physical meaning and possible practical applications
has been analyzed with a wealth of technical tools (see, e.g.,
\cite{GUS}, \cite{Mos4}).

It is natural to suppose that a spectral singularity
$\lambda_0\in(0,\infty)$ of $H_a$ is characterized by an
untypical behaviour of the resolvent $(H_a-\lambda{I})^{-1}$ in a
neighborhood of $\lambda_0$. This assumption leads to the following
definition: \emph{a positive number $\lambda_0$ is called
\emph{spectral singularity} of $H_a$ if there exists
$f\in{L_2(\mathbb{R})}$ such that the evaluation (\ref{ggg15}) does
not hold when non real $\lambda$ tends to $\lambda_0$}.

\begin{theorem}\label{ggg16}
Let $\lambda_0\in(0,\infty)$ and let there exist a sequence of
non-real $\lambda_n$ such that $\lambda_n\to\lambda_0$ and
$\lim_{n\to\infty}\widetilde{W}_{\lambda_n}=a\in\mathbb{C}\setminus\mathbb{R}$.
Then $\lambda_0$ is a spectral singularity of non-self-adjoint
operators $H_a$ and $H_{a^*}$.
\end{theorem}
\emph{Proof.} The inequality (\ref{ggg15}) is equivalent to the inequality
\begin{equation}\label{ggg15c}
\|\Xi(\lambda)f\|\leq\frac{C}{|Im\ \lambda|}\|f\|,
\end{equation}
where $\Xi(\lambda)$ is defined by (\ref{bbb9}). Moreover, it
follows from the proof of Theorem \ref{ggg21} that it is sufficient
to verify (\ref{ggg15c}) for $f=u_{\lambda^*}$ only.
By virtue of (\ref{ggg17}) and
the proof of Theorem \ref{ggg21},
\begin{equation}\label{ggg23}
\|\Xi(\lambda){u}_{\lambda^*}\|=\frac{\|
\gamma(\lambda)\gamma(\lambda^*)^\dag{u}_{\lambda^*}\|}{|a-\widetilde{W}_{\lambda}|}
=\frac{Im \ \widetilde{W}_\lambda}{Im \
\lambda}\frac{\|u_\lambda\|}{|a-\widetilde{W}_{\lambda}|}.
\end{equation}
It follows from (\ref{ggg22}) that
$\|u_\lambda\|=\|u_{\lambda^*}\|$. Replacing $\|u_\lambda\|$ by
$\|u_{\lambda^*}\|$ in (\ref{ggg23}) we rewrite (\ref{ggg15c}) in the
following equivalent form
  \begin{equation}\label{ggg15b}
\frac{|Im \
\widetilde{W}_\lambda|}{|a-\widetilde{W}_{\lambda}|}\leq{C}, \qquad
\lambda\in\mathbb{C}\setminus\mathbb{R}.
\end{equation}

If the condition of Theorem \ref{ggg16} is satisfied, then the
inequality (\ref{ggg15b}) cannot be true in neighborhood of
$\lambda_0$. Therefore, $\lambda_0$ should be a spectral singularity
of $H_a$. The same result holds for $H_{a^*}$ if we consider the
sequences $\lambda_n^*\to\lambda_0$,
 \ $W_{\lambda_n^*}={W_{\lambda_n}^*}\to{a^*}$ and take into account
that $H_a^\dag=H_{a^*}$.  \rule{2mm}{2mm}

\smallskip

If $\lambda=k^2$ with $k\in\mathbb{R}\setminus\{0\}$, then the
formula (\ref{ggg41}) allows one to define two functions
$u^\pm_\lambda$ corresponding to positive/negative values of $k$,
respectively. In this case, the formula
$$
\widetilde{W}^\pm_\lambda=[u^{\pm'}_\lambda]_s(0)-(q, u^\pm_\lambda)
=2ik\left(1+\frac{i}{k}\int_0^\infty{e^{iks}}q^{ev}(s)ds\right)-(q,
u^\pm_\lambda)
$$
($q^{ev}$ is the even part of $q$) gives two values of the Weyl-Titchmarsh
function $\widetilde{W}_\lambda$ on $(0,\infty)$.

The conditions imposed on $q$ guaranties that
$\widetilde{W}^\pm_\lambda$ are well-posed (i.e.
$\widetilde{W}^\pm_\lambda\not=\infty$).  Moreover,
the functions $\widetilde{W}^\pm_\lambda$ can be interpreted as
limits on $(0, \infty)$ of the holomorphic functions
$\widetilde{W}_\lambda$ considered on $\mathbb{C_{\pm}}$,
respectively. Taking the relation
$\widetilde{W}_\lambda^*=\widetilde{W}_{\lambda^*}$,
$\lambda\in\mathbb{C}\setminus[0,\infty)$ into account, we deduce
that $(\widetilde{W}_\lambda^+)^*=\widetilde{W}_{\lambda}^-$ for
$\lambda>0$. This relation and the definition of
$\widetilde{W}^\pm_\lambda$ imply that $u^+_\lambda$ and $u^-_\lambda$ are generalized
eigenfunctions of the operators $H_a$ and $H_{a^*}$, respectively
with $a=\widetilde{W}_\lambda^+$.

If $a=\widetilde{W}_\lambda^+$ is non-real, then, due to Theorem
\ref{ggg16}, $\lambda$  is a spectral singularity of the
non-self-adjoint operators $H_a$ and $H_{a^*}$. The corresponding
generalized eigenfunctions coincide with $u^+_\lambda$ and
$u^-_\lambda$. If $a=\widetilde{W}_\lambda^+$ is real, then the
evaluation (\ref{ggg15}) holds (since $H_a$ is self-adjoint) and
$\lambda$ cannot be a spectral singularity of $H_a$.

\section{Examples}\label{subsec4.3}

\subsection{Ordinary $\delta$-interaction}

This simplest case corresponds to $q=0$. The operators
$H_a=-\frac{d^2}{dx^2}$ have the domains:
$$
 \mathcal{D}(H_{a})=\left\{f\in{W}_2^2(\mathbb{R}\backslash\{0\}) \ : \
 \begin{array}{l}
 f(0-)=f(0+)\equiv{f(0)}  \vspace{2mm} \\
 f'(0+)-f'(0-)=af(0) \end{array} \right\}.
$$

The Weyl-Titchmarsh function has the form
$\widetilde{W}_\lambda=2ik=2i\sqrt{\lambda}$. There are no
exceptional points for operators $H_a$ because
$\widetilde{W}_\lambda'=i/\sqrt{\lambda}$ does not vanish on
$\mathbb{C}\setminus{[0,\infty)}$.

The limit functions $\widetilde{W}_\lambda^\pm=2ik$,  \
$k>{0}/k<{0}$ takes non-real values. Hence, the operators
$H_{\widetilde{W}_\lambda^+}$ and $H_{\widetilde{W}_\lambda^-}$ have
the spectral singularity $\lambda=k^2$.

The ordinary $\delta$-interaction are well-studied
\cite{deltainteraction}, \cite{Grod} and the evolution of spectral properties of
$H_a$ when $a$ runs $\mathbb{C}$ can be illustrated as follows:

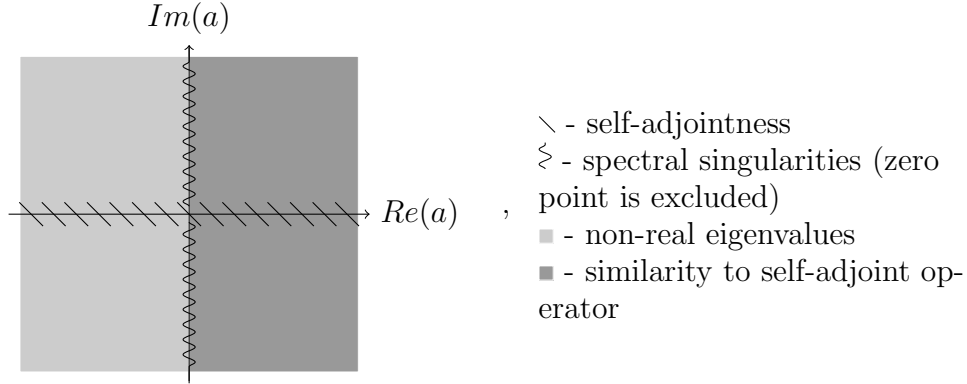
\begin{center}
\begin{tikzpicture}[scale=1.5]
  \filldraw[fill=black!20, draw=white] (-1.5cm, -1.4cm) rectangle (0, 1.4cm);
  \filldraw[fill=black!40, draw=white] (0, -1.4cm) rectangle (1.5cm, 1.4cm);

  \draw [->] (-1.6cm, 0) -- (1.6cm, 0) node[right] {$Re(a)$};
  \draw [->] (0, -1.5cm) -- (0, 1.5cm) node[above] {$Im(a)$};

  \foreach \x in {-1.4cm,-1.2cm,...,1.4cm}
    \draw (\x-3pt,3pt) -- (\x+3pt,-3pt);

  \draw [decorate,decoration={snake,amplitude=0.8mm,segment length=2mm,post length=1mm}]
        (0,2pt) -- (0,1.48cm);
  \draw [decorate,decoration={snake,amplitude=0.8mm,segment length=2mm,post length=1mm}]
        (0,-2pt) -- (0,-1.48cm);

\draw[xshift=3cm]
  node[left, text width=0.2cm]
    {
     ,
    }
  node[right,text width=6cm]
    {
      \begin{tikzpicture}
        \draw (-3pt,3pt) -- (3pt,-3pt);
      \end{tikzpicture} - self-adjointness\\

      \begin{tikzpicture}
        \draw [decorate,decoration={snake,amplitude=0.8mm,segment length=2mm,post length=0}]
        (0,0) -- (0,10pt);
      \end{tikzpicture} - spectral singularities (zero point is excluded)\\

      \begin{tikzpicture}
        \filldraw[fill=black!20, draw=white] (0.15cm,-0.15cm) rectangle (0.34cm, -0.34cm);
      \end{tikzpicture} - non-real eigenvalues\\

      \begin{tikzpicture}
        \filldraw[fill=black!40, draw=white] (0.15cm,-0.15cm) rectangle (0.34cm, -0.34cm);
      \end{tikzpicture} - similarity to self-adjoint operator\\
    };
\end{tikzpicture}
\end{center}

\smallskip

\subsection{The case of an odd function}

Let $q$ be an odd function. Then the Weyl-Titchmarsh function
$\widetilde{W}_\lambda$ takes especially simple form:
\begin{equation}\label{ggg61}
\widetilde{W}_\lambda=2ik-(q, u_\lambda)=2ik+(q, G\ast{q}), \qquad
\lambda=k^2, \quad k\in\mathbb{C}_+.
\end{equation}
The last equality in (\ref{ggg61}) follows from (\ref{ggg14}) since
$(G\ast{q})(0)=(G\ast{q^*})(0)=0$ for odd functions $q$, while the
second one is the consequence of  (\ref{au5}) and the fact that
$[u'_\lambda]_s(0)=2ik[1+(G\ast{q})(0)]=2ik$.

Let us consider, for simplicity, the odd function
$$
q(x)=Z\textsf{sign}(x)\chi_{[-\rho,\rho]}(x)=\left\{
\begin{array}{l}
Z, \quad 0\leq{x}\leq\rho \\
-Z, \quad -\rho\leq{x}<0 \\
0, \quad x\in\mathbb{R}\setminus[-\rho,\rho]
\end{array}\right. \quad Z\in\mathbb{C}, \quad \rho>0.
$$

The corresponding operators
$H_af=-\frac{d^2f}{dx^2}+f(0)Z\textsf{sign}(x)\chi_{[-\rho,\rho]}(x)$
with domains of definition
$$
\mathcal{D}(H_a)=\left\{f\in{W}_2^2(\mathbb{R}\backslash\{0\}) \ : \
\begin{array}{l}
 f(0-)=f(0+)\equiv{f(0)}  \vspace{2mm} \\
 f'(0+)-f'(0-)=af(0)+Z^*\int_{-\rho}^{\rho}\textsf{sign}(x)f(x)dx \end{array}
 \right\}
$$
have no positive eigenvalues (see Example \ref{ggg57}).

After the substitution of $q$ into (\ref{ggg61}) and elementary
calculations with the use of (\ref{ggg41}), we obtain the explicit
expression of the Weyl-Titchmarsh function
\begin{equation}\label{au6}
\widetilde{W}_\lambda=2ik-\frac{|Z|^2}{ik^3}\left[(e^{ik\rho}-2)^2+2ik\rho-1\right],
 \quad \lambda=k^2, \quad k\in\mathbb{C}_+.
\end{equation}

The limit functions $\widetilde{W}_\lambda^\pm$ are determined by
(\ref{au6}) for   $k>{0}$ and $k<{0}$, respectively. It is easy to
check that the imaginary part of $\widetilde{W}_\lambda^\pm$:
$$
\textsf{Im} \
\widetilde{W}_\lambda^\pm=2k+\frac{|Z|^2}{k^3}(2\cos^2{k\rho}-4\cos{k\rho}+2)
$$
do not vanish when $k$ runs $\mathbb{R}\setminus\{0\}$. Hence, any positive
$\lambda$ turns out to be a spectral singularity for some operators $H_a$. Namely,
the operators $H_a$ and $H_{a^*}$  with $a={\widetilde{W}_\lambda^+}$
will have the spectral singularity $\lambda$.

\smallskip

\subsection{The case of even function $q=ce^{-\mu|x|}$ \ ($\mu>0$)}

The corresponding operators
$H_af=-\frac{d^2f}{dx^2}+f(0)ce^{-\mu|x|}$ have the domains
$$
\mathcal{D}(H_a)=\left\{f\in{W}_2^2(\mathbb{R}\backslash\{0\}) \ : \  \begin{array}{l}
 f(0-)=f(0+)\equiv{f(0)}  \vspace{2mm} \\
 f'(0+)-f'(0-)=af(0)+c^*\int_{\mathbb{R}}e^{-\mu|x|}f(x)dx \end{array} \right\}.
$$

The eigenfunctions $u_\lambda$ (see (\ref{ggg41})) are given by the expression
\begin{equation}\label{au8}
u_\lambda=\left(1-\frac{c}{\mu^2+\lambda}\right)e^{ik|x|}+\frac{q(x)}{\mu^2+\lambda},
\qquad \lambda=k^2.
\end{equation}

The Weyl-Titchmarsh function
\begin{equation}\label{ggg61b}
\widetilde{W}_\lambda=2ik-(q, u_\lambda)=2ik-\frac{4\textsf{Re} \
c}{\mu-ik}+\frac{\|q\|^2}{(\mu-ik)^2}
\end{equation}
is defined on $\mathbb{C}\setminus[0,\infty)$ and its limit functions
$\widetilde{W}_\lambda^\pm$ are determined by (\ref{ggg61b})
with $k>{0}$ and $k<{0}$, respectively.

Each $\lambda\in\mathbb{C}\setminus[0,\infty)$ is an eigenvalue of
the operator $H_a$ with $a=\widetilde{W}_\lambda$ and the
corresponding eigenfunction is given by (\ref{au8}).

It follows from (\ref{au8}) that a positive eigenvalue $\lambda$
exists for some operator $H_a$ \emph{if and only if} $c\geq\mu^2$.
In this case, $\lambda=c-\mu^2$, the corresponding eigenfunction
$u_\lambda$ coincides with $\frac{q(x)}{\mu^2+\lambda}=e^{-\mu|x|}$
and $u_\lambda$ an eigenfunction of a self-adjoint operator $H_a$
with $a=\widetilde{W}_\lambda^{\pm}=-3\mu-\frac{\lambda}{\mu}$.

Let us assume for the simplicity that $c\in{i\mathbb{R}}$ and $\|q\|^2=\frac{|c|^2}{\mu}=1$. Then
\begin{equation}\label{ggg61c}
\widetilde{W}_\lambda=2ik+\frac{1}{(\mu-ik)^2}=2i\sqrt{\lambda}+\frac{1}{(\mu-i\sqrt{\lambda})^2}.
\end{equation}

If $k$ is real in (\ref{ggg61c}), then the imaginary part of $\widetilde{W}_\lambda^\pm$:
$$
\textsf{Im} \ \widetilde{W}_\lambda^\pm=2k+\frac{2k\mu}{|\mu-ik|^2}
$$
does not vanish when $\lambda=k^2\in(0,\infty)$.
Hence, any positive $\lambda$ is a spectral singularity of operators $H_a$ and $H_{a^*}$ with
$a={\widetilde{W}_\lambda^+}$.

It follows from (\ref{ggg61c}) that
$$
\widetilde{W}_\lambda'=\frac{i}{k}\left[1+\frac{1}{(\mu-ik)^3}\right]
=\frac{i}{\sqrt{\lambda}}\left[1+\frac{1}{(\mu-i\sqrt{\lambda})^3}\right].
$$
Therefore, $\widetilde{W}_\lambda'=0$  for certain
$\lambda\in\mathbb{C}\setminus{[0,\infty)}$ if and only if
$(\mu-ik)^3=-1$ for $k\in\mathbb{C}_+$. The latter equation has two
required solutions
$$
k_0=\frac{\sqrt{3}}{2}+i(\frac{1}{2}-\mu), \qquad k_1=-k_0^*
$$
when $0<\mu<\frac{1}{2}$. By virtue of Theorem \ref{ggg21},
$\lambda_0=k_0^2$ is an exceptional point of the operator $H_a$ with
$$
a=\widetilde{W}_{\lambda_1}
=2ik_0+\frac{1}{(\mu-ik_0)^2}=2ik_0+\frac{\mu-ik_0}{(\mu-ik_0)^3}=3ik_0-\mu,
$$
while $\lambda_1=k_1^2=\lambda_0^*$ will be an exceptional point of
its adjoint $H_{a^*}=H_{a}^\dag$, cf. Corollary \ref{au10}.

The obtained result shows that the existence of exceptional points
for some operators from the collection $\{H_a\}_{a\in\mathbb{C}}$
depends on the behaviour of the function $q(x)=ce^{-\mu|x|}$.  If
$q(x)$ decrease (relatively) slowly on $\infty$ (the case
$0<\mu<\frac{1}{2}$) then exist two operators $H_{a}$ and $H_a^\dag$
with exceptional points $\lambda_0$ and $\lambda_0^*$, respectively.

\section{Summary\label{summary}}

Although the knowledge of the merits of the pseudo-Hermitian
representation of observables (and, in particular, of Hamiltonians)
in quantum theory dates back to the middle of the last
century, its applicability still remains restricted,
mainly due to the presence and emergence of multiple technical
obstacles \cite{MZbook}. In the present paper we paid attention to
the possibilities of circumventing the obstacles via introduction of
interactions which combined the exact solvability feature of the
traditional point interactions with the necessity of extension of
the latter class of local potentials to some maximally friendly
nonlocal generalizations.

For the sake of a reasonable length of our paper we only considered
a subset of the eligible candidates for the interaction and we also
did not pay any explicit attention to the possible connection of our
models with physics and with the possible experimental realizations
of the systems.  This enabled us to pay more attention to the usually neglected
mathematical features of the models and to the explicit description
of the qualitative differences between the self-adjoint and
non-selfadjoint choices and/or between the local and nonlocal
versions and special cases of the Hamiltonians.

We would like to emphasize the importance of our present successful
transition from the traditional study of finite-matrix models (i.e.,
of the simplified, difference Schr\"{o}dinger equations as sampled,
e.g., in \cite{oboje}) to the full-fledged differential
operators (albeit with the mere ultralocal-distribution
interactions). Obviously, such a step still remains to be followed
by several future resolutions of challenges incorporating, first of
all, the construction of the physical inner products, etc.

In a way inspired by the older developments in self-adjoint context
\cite{Nizhnik} we found a key to the technical new results in the
use of the language of the formalism of boundary triplets. We
managed to demonstrate that even after a restriction of our
attention to the first nontrivial class of one point nonlocal
interactions the wealth of the spectral properties of the models
remains satisfactorily rich involving not only the usual
regularities/anomalies in the discrete spectra but, equally well,
also the advanced (and, in the finite-dimensional models,
inaccessible) features of the presence of the exceptional points and
of the spectral singularities.

Naturally, we expect that the set of the present results will be
complemented, in some not too remote future, not only by the similar
rigorous coverage of the more general nonlocal interactions (and of
the related enhanced flexibility, say, in the quantum spectral
design) but also by the development of some parallels to the success
of transfer of the applicability of the manifestly non-selfadjoint
models in the scattering dynamical regime, with a particular
emphasis upon the possible restoration of the unitarity of the S
matrix (in this direction our future plans will be inspired by the
encouraging success of Ref.~\cite{treci} in the analysis of certain
local point-interaction predecessors of our present models).

\section{Appendix: Boundary triplets}

Let $S_{\min}$ be a closed symmetric (densely defined) operator
 in a Hilbert space $\mathfrak{H}$ with inner product
$(\cdot,\cdot)$. Denote
$S_{max}=S_{\min}^\dagger$. Obviously, $S_{\min}\subset{S_{max}}$.

A triplet $(\mathcal{H}, \Gamma_0, \Gamma_1)$, where $\mathcal{H}$ is an
auxiliary Hilbert space and $\Gamma_0$, $\Gamma_1$ are linear
mappings of $\mathcal{D}(S_{max})$ into $\mathcal{H}$, is called a
\emph{boundary triplet of} $S_{max}$ if the Green identity
$$
(S_{max}f, g)-(f, S_{max}g)=(\Gamma_1f, \Gamma_0g)_{\mathcal{H}}-(\Gamma_0f,
\Gamma_1g)_{\mathcal{H}}, \quad  f, g\in\mathcal{D}(S_{max})
$$
is satisfied and the map $(\Gamma_0,
\Gamma_1):\mathcal{D}(S_{max})\to\mathcal{H}\oplus\mathcal{H}$ is
surjective.

The symmetric operator $S_{min}$ is the restriction of $S_{max}$
onto $\mathcal{D}(S_{min})=\{f\in\mathcal{D}(S_{max}) \ : \
\Gamma_0f=\Gamma_1f=0\}$. The defect indices of $S_{min}$ coincides
with the dimension of  ${\mathcal{H}}$. Boundary triplets of
$S_{max}$ are not determined uniquely and they exist only in the
case where the symmetric operator $S_{min}$ has self-adjoint
extensions\footnote{see \cite{KK} for various generalization of
boundary triplets}.

Let $(\mathcal{H}, \Gamma_0, \Gamma_1)$ be a boundary triplet of
$S_{max}$.  Then the operator
$$
H_\infty=S_{max}\upharpoonright_{\mathcal{D}(H_\infty)}, \qquad \mathcal{D}(H_\infty)
=\{f\in\mathcal{D}(S_{max}) \ : \ \Gamma_0f=0\}
$$
is a self-adjoint extension of $S_{min}$.

The Weyl-Titchmarsh function $W_\lambda$ associated to the boundary
triplet $(\mathcal{H}, \Gamma_0, \Gamma_1)$ is defined for all
$\lambda\in\rho(H_\infty)$ \cite{DM}:
$$
W_\lambda\Gamma_0f_\lambda=\Gamma_1f_\lambda, \qquad
\forall{f}_\lambda\in\ker(S_{max}-\lambda{I}).
$$

The operator valued function $W_\lambda$ is holomorphic on
$\rho(H_\infty)$ and the adjoint of the operator $W_\lambda$ in
$\mathcal{H}$ coincides with $W_{\lambda^*}$.

Let $f_\lambda\in\ker(S_{max}-\lambda{I})$. It follows from the
Green identity that
\begin{equation}\label{ggg8}
(Im \ \lambda)\|f_\lambda\|^2=(\Gamma_0f_\lambda, (Im \
W_\lambda)\Gamma_0f_\lambda),
 \quad \mbox{where} \quad  Im \ W_\lambda=\frac{W_\lambda-W_\lambda^\dag}{2i}.
\end{equation}
Therefore, $(Im \ \lambda)(Im \ W_\lambda)>0$ for non-real
$\lambda$. The latter means that  $W_\lambda$ is a Herglotz
(Nevanlinna) function \cite{GT}.

Let ${\mathbf{T}}$ be a bounded operator in the auxiliary
Hilbert space $\mathcal{H}$. The operator
$$
H_{\mathbf{T}}=S_{max}\upharpoonright_{\mathcal{D}(H_{\mathbf{T}})},
\qquad \mathcal{D}(H_\mathbf{T}) =\{f\in\mathcal{D}(S_{max}) \ : \
(\mathbf{T}\Gamma_0-\Gamma_1)f=0 \}
$$
is a proper extension of $S_{min}$ (i.e., $S_{min}\subset{H_{\mathbf{T}}}\subset{S_{max}}$). Moreover, the
adjoint operator $H_{\mathbf{T}}^\dagger$ is also a proper
extension and $H_{\mathbf{T}}^\dagger=H_{\mathbf{T}^\dagger}$,  where
$\mathbf{T}^\dagger$ is the adjoint operator of $\mathbf{T}$ in the
auxiliary space $\mathcal{H}$. Hence, the self-adjointness of
unbounded operator $H_{\mathbf{T}}$ in $\mathfrak{H}$  is equivalent to the
self-adjointness of bounded operator $\mathbf{T}$ in the auxiliary space
$\mathcal{H}$.

The spectrum of $H_{\mathbf{T}}$ is described in terms of
${\mathbf{T}}$ and $W_\lambda$. Namely \cite{DM},
$\lambda\in\rho(H_\infty)$ belongs to the point
$\sigma_p(H_{\mathbf{T}})$, to the residual
$\sigma_r(H_{\mathbf{T}})$, and to the continuous
$\sigma_c(H_{\mathbf{T}})$ parts of the spectrum of $H_\infty$ if
and only if $0$ belongs to the same parts of spectrum of
$\mathbf{T}-W_\lambda$, i.e., if
$0\in\sigma_\alpha(\mathbf{T}-W_\lambda)$, \ $\alpha\in\{p,r,c\}$.

For each $\lambda\in\rho(H_\infty)$, the operator $\Gamma_0$ is a bijective mapping of the subspace
$\ker(S_{max}-\lambda{I})$ onto $\mathcal{H}$. Its bounded inverse
$$
\gamma(\lambda)=(\Gamma_0\upharpoonright_{\ker(S_{max}-\lambda{I})})^{-1}
: \mathcal{H} \to \ker(S_{max}-\lambda{I})
$$
is called the $\gamma$-field associated with $(\mathcal{H}, \Gamma_0, \Gamma_1)$.

The $\gamma$-field $\gamma(\cdot)$ is a holomorphic operator-valued
function on $\rho(H_\infty)$ and \cite[Prop. 14.14, 14.15]{Schm}
\begin{equation}\label{ggg11}
\gamma(\lambda^*)^\dag=\Gamma_1(H_{\infty}-\lambda{I})^{-1},
 \qquad \frac{d}{d\lambda}W_\lambda=\gamma(\lambda^*)^\dag\gamma(\lambda)
\end{equation}
where the adjoint operator $\gamma(\lambda^*)^\dag$ maps
$\ker(S_{max}-\lambda^*{I})$ into $\mathcal{H}$.

For any $\lambda\in\rho(H_\infty)\cap\rho(H_{\mathbf{T}})$, the
Krein-Naimark resolvent formula
\begin{equation}\label{ggg9}
(H_{\mathbf{T}}-\lambda{I})^{-1}-(H_\infty-\lambda{I})^{-1}
=\gamma(\lambda)(\mathbf{T}-W_\lambda)^{-1}\gamma(\lambda^*)^\dag
\end{equation}
holds \cite[Theorem 14.18]{Schm}.

Let us assume for simplicity that the auxiliary space $\mathcal{H}$
is \emph{finite-dimensional}, i.e., $\dim\mathcal{H}=m$ and the
spectrum of $H_{\infty}$ is \emph{purely continuous}.
Then, the continuous spectrum of each $H_{\mathbf{T}}$
coincides with $\sigma(H_\infty)$ and only eigenvalues of
$H_{\mathbf{T}}$ may appear in
$\mathbb{C}\setminus\sigma(H_\infty)$ (since $H_{\mathbf{T}}$ are
finite rank perturbations of the self-adjoint operator $H_\infty$).
Without loss of generality, we may assume that $\mathcal{H}=\mathbb{C}^m$.  In this
case, the operator $\mathbf{T}$ and the Weyl-Titchmarsh function
$W_\lambda$  can be replaced by $m\times{m}$--matrices and
$\lambda\in\mathbb{C}\setminus\sigma(H_{\infty})$ is an eigenvalue
of $H_{\mathbf{T}}$ if and only if
$\det(\mathbf{T}-W_{\lambda})=0.$ The geometric multiplicity of an
eigenvalue $\lambda$ coincides with
$m-\mbox{rank}(\mathbf{T}-W_{\lambda})$.

In our presentation we assume that $\sigma(H_{\mathbf{T}})\not=\mathbb{C}$.
Then, the presence of an eigenvalue
$\lambda_0\in\mathbb{C}\setminus\sigma(H_{\infty})$ of
$H_{\mathbf{T}}$ can be characterized as follows: $\lambda_0$ should
be a zero of finite-type \cite[Definition 3.1]{Jussi} of the
matrix-valued holomorphic function $\mathbf{T}-W_\lambda$.

According to the Fredholm theorem \cite[Thm. VI.14]{RS1},
$(\mathbf{T}-W_\lambda)^{-1}$ is holomorphic on the punctured disk
$D(\lambda_0; \epsilon_0)=\{\lambda\in\mathbb{C} \ | \
0<|\lambda-\lambda_0|<{\varepsilon_0}\}$ for some $0<\varepsilon_0$
sufficiently small. In this case, we may
define\footnote{\cite[Definition 4.2]{Jussi}} the index of
$\mathbf{T}-W_\lambda$ with respect to the counterclockwise oriented
circle $C(\lambda_0; \varepsilon)=\{\lambda\in\mathbb{C} \ | \
|\lambda-\lambda_0|={\varepsilon}\}$:
\begin{equation}\label{ggg3}
\mbox{ind}_{C(\lambda_0; \varepsilon)}(\mathbf{T}-W_\lambda)
=\mbox{tr}_{\mathbb{C}^m}\oint_{C(\lambda_0; \varepsilon)}
 {W_\xi'}(W_\xi-\mathbf{T})^{-1}d\xi, \quad 0<\varepsilon<\varepsilon_{0}.
\end{equation}

By virtue of \cite[Theorem 6.4]{Jussi} the algebraic multiplicity of
the eigenvalue $\lambda_0$ of $H_{\mathbf{T}}$ coincides with
$\mbox{ind}_{C(\lambda_0; \varepsilon)}(\mathbf{T}-W_\lambda)$. The
latter quantity is also the algebraic multiplicity of the zero of
$\mathbf{T}-W_\lambda$ at $\lambda_0$.


\begin{thebibliography}{}
\bibitem{CGM}
Caliceti, E., Graffi, S., Maioli, M.:
Perturbation theory of odd
anharmonic oscillators. Commun. Math. Phys. {\bf 75}, 51-–66 (1980) 

\bibitem{BG} Buslaev, V., Grecchi, V.: Equivalence of unstable anharmonic oscillators and double wells, J. Phys. A Math. Gen. 
{\bf 26}, 5541--5549 (1993)

\bibitem{BB} Bender, C. M., Boettcher, S.: 
 S. Real spectra in non-Hermitian Hamiltonians having PT symmetry.
Phys. Rev. Lett.  {\bf 80},  5243--5246 (1998)

\bibitem{book}
Bagarello F et al, Eds,:  Non-Selfadjoint Operators in Quantum
Physics: Mathematical Aspects. Wiley, Hoboken (2015)

\bibitem{Dieudonne}
Dieudonn\'{e}, J.:  Quasi-Hermitian operators. Proc. Int. Symp. Lin. Spaces, pp. 115--122, Pergamon, Oxford (1961)

\bibitem{Jones}
 Jones, H.F.: Scattering from localized non-Hermitian potentials. Phys. Rev. D
 {\bf 76}, 125003 (2007)

\bibitem{scatt}
 Znojil, M.: Scattering theory with localized non-Hermiticities.
Phys. Rev. D {\bf 78} 025026 (2008)

\bibitem{Nizhnik}
Albeverio, S., Nizhnik, L.: Schr\"{o}dinger operators with nonlocal point interactions. J. Math. Anal. Appl. {\bf 332}  884--895 (2007)

\bibitem{Nizhnik1}
Albeverio, S., Nizhnik, L.: Schr\"{o}dinger operators with nonlocal potentials. Methods Funct. Anal. Topology. {\bf 19}(3) 199--210 (2013)

\bibitem{Nizhnik1b}
Brasche, J. and Nizhnik, L. P.: One-dimensional Schrodinger operators with general point interactions. Methods Funct. Anal. Topology. {\bf 19}(1), 4--15 (2013)

\bibitem{Nizhnik2}
Albeverio S, Hryniv R and Nizhnik L.:  Inverse spectral problems for nonlocal Sturm-Liouville
operators. Inverse problems. {\bf 23} 523--536 (2007)

\bibitem{Nizhnik3}
Nizhnik, L.P.: Inverse nonlocal Sturm-Liouville problem. Inverse Problems.  {\bf 26} 125006--125015 (2010)

\bibitem{Nizhnik4}
Nizhnik, L.P.: Inverse spectral nonlocal problem for the first order ordinary differential equation. Tamkang Journal of Mathematics. {\bf 42}(3), 385--394 (2011)

\bibitem{Carl}
Bender, C.M.: Making sense of non-Hermitian Hamiltonians. Reports on Progress in Physics. {\bf 70} 947--1018 (2007)

\bibitem{Mos}
Mostafazadeh, A.: Pseudo-Hermitian representation of quantum mechanics. Int. J. Geom. Meth. Mod. Phys. {\bf 7}(7) 1191-1306 (2010)

\bibitem{ZN}
Znojil, M.: Cryptohermitian Picture of Scattering Using Quasilocal Metric Operators. Symmetry, Integrability and Geometry: Methods and Applications (SIGMA). {\bf 5}, 85--106 (2009) 

\bibitem{nonlocal}
 Albeverio, S.,  Kuzhel, S.: One-dimensional Schr\"{o}dinger operators with $\mathcal{P}$-symmetric zero-range potentials. J. Phys. A: Math. Gen. {\bf 38}, 4975--4988 (2005)
 
\bibitem{Mos2} 
Mostafazadeh, A.: Spectral singularities of a general point interaction. J. Phys. A: Math. Gen. {\bf 44}, 375302--375311 (2011)

\bibitem{ZN3}
Znojil, M., Jakubsky, V.: Solvability and PT-symmetry in a double-well model with point interactions. J. Phys. A {\bf 38} 5041-5056 (2005)

\bibitem{AK_Albeverio0}
  Albeverio, S.,  Gesztesy, F., H{\o}egh-Krohn, R., Holden, H.: 
  Solvable Models in Quantum Mechanics. 2nd ed. with an Appendix by Exner P. AMS, Providence, Chelsea
Publishing (2005) 

\bibitem{deltainteraction}
 Mostafazadeh, A.: Delta-Function Potential with a Complex Coupling. J. Phys. A: Math. Gen. {\bf 39}, 13495-13506 (2006)

\bibitem{Grod}
Grod, A., Kuzhel, S.: Schr\"{o}dinger operators with non-symmetric zero-range potentials. Methods of Functional Analysis and Topology. {\bf 20}(1), 34-49 (2014)

\bibitem{Bernhdt}  
Behrndt, J.,  Langer, M.:  On the adjoint of a symmetric operator.  J. Lond. Math. Soc. {\bf 82} 563–-580 (2010)

\bibitem{Schm} Schm\"{u}dgen, K.:  Unbounded Self-adjoint Operators on Hilbert space.
Berlin, Springer (2012)

\bibitem{WOL}  Baumg\"{a}rtel, H.: Analytic Perturbation Theory for Matrix and Operators.
Basel, Birkh\"{a}user (1985)

\bibitem{Naim} Naimark, M.A.:   Investigation of the spectrum and the expansion in eigenfunctions of a non-selfadjoint differential operator of the second order on a semi-axis. Amer. Math. Soc. Transl.(2) {\bf 16}, 103--193 (1960)

\bibitem{GUS}  
Guseinov, G.Sh.: On the concept of spectral singularities. Pramana J. Phys. {\bf 73}, 587--603  (2009)

 \bibitem{Mos4} 
 Mostafazadeh, A.: Physics of Spectral Singularities, Geometric Methods in Physics, Trends in Mathematics 145-165, Springer, Cham (2015)
 
 \bibitem{MZbook}
 Znojil, M.: Non-selfadjoint operators in quantum physics: ideas, people and trends.  in Ref. \cite{book}, pp. 7 - 58.

\bibitem{oboje}
Znojil, M.: PT-symmetric model with an interplay between
kinematical and dynamical non-localities.
J. Phys. A: Math. Theor. {\bf 48},  195303-195319
(2015)

\bibitem{treci}
Cojuhari, P.A., Grod, A., Kuzhel, S.:   On $S$-matrix of Schr\"{o}dinger operators with non-symmetric
zero-range potentials. J. Phys. A: Math. Theor. {\bf 47}, 315201-315219 (2014)

\bibitem{KK}  Kuzhel, A., Kuzhel, S.: Regular Extensions of Hermitian Operators
Utrecht, VSP (1998)

\bibitem{DM} Derkach, V., Malamud, M.: Generalized resolvents and the boundary value problems for Hermitian operators with gaps. 
J. Funct. Anal. {\bf 95}, 1-95 (1991).

\bibitem{GT}  Gesztesy, F.,  Tsekanowskii, E.: On matrix-valued Herglotz functions. Math. Nachrichten {\bf 218},  
61-138 (2000).

\bibitem{Jussi} Behrndt, J., Gesztesy, F.,  Holden, H., Nichols, R.: arXiv:1512.06962v2 (2016)

\bibitem{RS1}  Reed, M., Simon, B.: Methods of Modern Mathematical Physics I. Functional Analysis.  New York, Academic Press
(1980)
\end{thebibliography}
\end{document}